\documentclass[aps,prc,showpacs,twocolumn,nofootinbib,groupedaddress,preprintnumbers]{revtex4-1}

\usepackage{calligra}
\usepackage{longtable}
\usepackage{graphicx}
\usepackage{dcolumn}
\usepackage{bm}
\usepackage{amsmath,amssymb}
\usepackage{color}

\newcommand{\re}{\text{Re}}

\begin{document}

\preprint{YITP-15-53}


\title{Structure of $\Lambda(1405)$ and construction of $\bar{K}N$ local potential \\ based on chiral SU(3) dynamics}


\author{Kenta~Miyahara}
\email[]{miyahara@ruby.scphys.kyoto-u.ac.jp}
\affiliation{Department of Physics, Graduate School of Science, Kyoto University, Kyoto 606-8502, Japan}
\author{Tetsuo~Hyodo}
\affiliation{Yukawa Institute for Theoretical Physics, Kyoto University, Kyoto 606-8502, Japan}


\date{\today}

\begin{abstract}                           
We develop the single-channel local potential for the $\bar{K}N$ system, which is applicable to quantitative studies of $\bar{K}$ bound states in nuclei. Because the high precision measurement of the kaonic hydrogen by SIDDHARTA reduces the uncertainty of the $\bar{K}N$ amplitude below the $\bar{K}N$ threshold, the local potential should be calibrated in a wide energy region. We establish a new method to construct the local potential focusing on the behavior of the scattering amplitude in the complex energy plane. Applying this method, we construct the $\bar{K}N$ potential based on the chiral coupled-channel approach with the SIDDHARTA constraint. The wave function from the new potential indicates the $\bar{K}N$ molecular structure of $\Lambda(1405)$. 
\end{abstract}




\maketitle

\section{Introduction}  \label{sec:intro}  

The multi-nucleon systems with an antikaon draw significant attention in hadron and nuclear physics. It is considered that the strong attraction in the $\bar{K}N$ channel leads to various interesting phenomena. The simplest example is the $\Lambda(1405)$ resonance as a $\bar{K}N$ quasi-bound state~\cite{Dalitz:1959dn,Dalitz:1960du}. The difficulty of the description of $\Lambda(1405)$ by the three-quark picture in the constituent quark model~\cite{Isgur:1978xj} is compatible with the interpretation as the quasi-bound state of the $\bar{K}N$ system slightly below the threshold. 
In the early days, $\Lambda(1405)$ in the $\bar{K}N$ scattering amplitude was analyzed  with $K$-matrix approaches and dispersion relations~\cite{Kim:1965zzd,Kim:1967zze,Martin:1969fn,Martin:1969bk,Martin:1969yz,Martin:1976xc,Martin:1980qe}.
Later, coupled-channel approaches with chiral SU(3) symmetry~\cite{Kaiser:1995eg,Oset:1998it,Oller:2000fj,Lutz:2001yb,Hyodo:2011ur} have been applied, and confirmed the $\bar{K}N$ quasi-bound picture of $\Lambda(1405)$. 
In addition, a recent lattice QCD analysis supports this picture based on the vanishing of the strange magnetic form factor~\cite{Hall:2014uca}. The $\bar{K}N$ molecule picture indicates that the spacial structure of $\Lambda(1405)$ is relatively larger than the usual hadronic scale, which is shown by several approaches~\cite{Yamazaki:2007cs,Sekihara:2008qk,Sekihara:2010uz,Sekihara:2012xp,Dote:2014ema}. Experimentally, the $\Lambda(1405)$ signal in the $\pi\Sigma$ spectrum has been studied with various reactions~\cite{Niiyama:2008rt,Agakishiev:2012xk,Moriya:2013eb,Moriya:2013hwg,Moriya:2014kpv}.

Another interesting example is the antikaon bound states in nuclei, the $\bar{K}$-nuclei~\cite{Nogami:1963xqa,Akaishi:2002bg,Yamazaki:2002uh}. Because of the strong $\bar{K}N$ attraction, the $\bar{K}$-nuclei may exhibit the qualitatively different structure from the normal nuclei. Experimentally, there have been some claims for the evidence of the $\bar{K}NN$ state~\cite{Agnello:2005qj,Bendiscioli:2007zza,Yamazaki:2010mu,Ichikawa:2014ydh}. For instance, J-PARC E27 experiment has reported a broad enhancement in the proton coincidence missing mass spectra in the $d(\pi^+,K^+)$ reaction at 1.69 GeV/$c$~\cite{Ichikawa:2014ydh}. 
However, we have to note that quantitative results of these experiments are not consistent with each other. Furthermore there are experiments which have found no such quasi-bound structure~\cite{Tokiyasu:2013mwa,Hashimoto:2014cri,Agakishiev:2014dha}. To draw a definite conclusion, further studies are needed. Theoretically, the rigorous three-body calculations of the $\bar{K}NN$ system have recently been performed~\cite{Shevchenko:2006xy,Shevchenko:2007ke,Yamazaki:2007cs,Ikeda:2007nz,Dote:2008in,Dote:2008hw,Ikeda:2010tk,Barnea:2012qa,Dote:2014via}. All calculations have obtained the qualitatively consistent result that the $\bar{K}NN$ system is bound between the $\bar{K}NN$ and $\pi\Sigma N$ thresholds. However, the quantitative predictions of the mass and the width are substantially different from each other and are not consistent with the experimentally reported values. In this way, the quantitatively conclusive result of the $\bar{K}NN$ system has not been achieved.

It is the $\bar{K}N$ interaction below the threshold that is essential for the calculations of $\bar{K}$-nuclear systems. However, the subthreshold region cannot be directly accessible by experiments, so we have to extrapolate the scattering amplitude constrained by the experimental data above the $\bar{K}N$ threshold. Previous studies of the $\bar{K}$-nuclei have suffered from the large uncertainty, mainly because the experimental data has not been sufficient to constrain the subthreshold amplitude. Recently, the SIDDHARTA collaboration has measured the precise energy-level shift of the kaonic hydrogen~\cite{Bazzi:2011zj,Bazzi:2012eq}. This data is related to the $K^-p$ scattering length \cite{Meissner:2004jr}, which quantitatively constrains the scattering amplitude at the $\bar{K}N$ threshold. This reduces the uncertainty of the amplitude below the $\bar{K}N$ threshold significantly~\cite{Ikeda:2011pi,Ikeda:2012au}. For a reliable prediction of $\Lambda(1405)$ and the $\bar{K}$-nuclei, the constraint from the SIDDHARTA data should be taken into account.

The base for the few-body calculations is the two-body hadron interaction. Historically, the hadron interaction has been constructed phenomenologically. In the case of the nuclear force, phenomenological interactions are quite successful in reproducing the experimental data with the precision of $\chi^2/{\rm d.o.f}\simeq 1$~\cite{Wiringa:1994wb,Machleidt:2000ge}. Though the phenomenological interactions have been successfully applied to various few-body systems, the direct connection to QCD is not obvious. The first principle calculation of QCD is the lattice simulation which provides the promising approach to the hadron potentials~\cite{Aoki:2012tk}. However, the nuclear force in the realistic set-up is yet to be constructed. Another approach is based on chiral perturbation theory which is the effective field theory of QCD with chiral symmetry being the guiding principle~ \cite{Epelbaum:2008ga,Machleidt:2011zz}. 
In this approach, the potential can be systematically improved with the higher order contributions. In the state-of-the-art calculations, it is possible to construct the nuclear force as precise as the phenomenological ones. 

In this work, we construct the $\bar{K}N$ potential using chiral unitary approach which is based on chiral perturbation theory and unitarity of the scattering amplitude. Thanks to the systematic improvement, the low energy $K^{-}p$ total cross sections, threshold branching ratios, and the SIDDHARTA data are well reproduced with an accuracy of $\chi^2/{\rm d.o.f}\simeq 1$~\cite{Ikeda:2011pi,Ikeda:2012au}. The potential is constructed in the local form in the coordinate space for the convenience of the applications to few-body calculations. In contrast to the nuclear force, the $\bar{K}N$ potential cannot be directly obtained in chiral perturbation theory 
which does not contain the long range meson exchange processes. 
We therefore construct the potential so as to reproduce the scattering amplitude from chiral unitary approach on the real energy axis following Ref.~\cite{Hyodo:2007jq}. Given that the uncertainty of the subthreshold amplitude is reduced by the SIDDHARTA constraint, we have to establish the construction procedure with the high precision in the wide energy region. Moreover, to analyze the structure of $\Lambda(1405)$, the precision in the complex energy plane is necessary. In this way, we construct the reliable $\bar{K}N$ potential applicable for the quantitative calculations. 

In Sec.~\ref{sec:formulation}, we briefly introduce chiral unitary approach for the $\bar{K}N$ scattering, and  the framework to construct the hadron local potential from this approach. 
In Sec.~\ref{sec:precisepot}, we examine the construction procedure to reproduce the original amplitude even in the complex energy plane with a simple model as an example. The new construction procedure of the hadron potential is applied to the $\bar{K}N$ amplitude with the SIDDHARTA constraint in Sec.~\ref{sec:application}, leading to the reliable $\bar{K}N$ potential.  Using this new $\bar{K}N$ potential, we investigate the spatial structure of $\Lambda(1405)$. The last section is devoted to the summary of this work.

\section{Formulation} \label{sec:formulation}  

\subsection{Chiral SU(3) dynamics for $\bar{K}N$ scattering}
To describe the $\bar{K}N$ scattering, it is mandatory to consider the channel coupling with the lower energy $\pi\Sigma$ state and the existence of the $\Lambda(1405)$ resonance below the threshold. Here we utilize the nonperturbative coupled-channel framework called chiral unitary approach \cite{Kaiser:1995eg,Oset:1998it,Oller:2000fj,Lutz:2001yb,Hyodo:2011ur} which is based on the resummation of the interaction terms derived from chiral perturbation theory. The $s$-wave meson-baryon scattering amplitude $T_{ij}(\sqrt{s})$ at the total center of mass energy $\sqrt{s}$ is
\begin{align}
T_{ij}(\sqrt{s}) &= V_{ij}(\sqrt{s}) + V_{ik}(\sqrt{s})G_{k}(\sqrt{s})T_{kj}(\sqrt{s})    \label{eq:BetheSal} \\
&= \left[ ({V(\sqrt{s})}^{-1}-G(\sqrt{s}))^{-1} \right]_{ij},  \notag
\end{align}
where $V_{ij}$ and $G_i$ represent the meson-baryon interaction kernel derived from chiral perturbation theory and the loop function, respectively with the meson-baryon channel indices being denoted by $i,j$. There are four meson-baryon channels with isospin $I=0$, $\bar{K}N$, $\pi\Sigma$, $\eta\Lambda$ and $K\Xi$ corresponding to $i=1,2,3$ and 4, respectively. 
The interaction kernel $V_{ij}$ is systematically obtained in chiral perturbation theory, where the leading contribution is given by the Weinberg-Tomozawa term. 
Systematic improvement with higher order correction has been discussed in Ref.~\cite{Kaiser:1995eg,Lutz:2001yb,Borasoy:2005fq,Borasoy:2005ie,Borasoy:2006sr,Borasoy:2004kk}.
Recently, the refined calculations for the $S=-1$ sector including the next-to-leading order terms~\cite{Ikeda:2011pi,Ikeda:2012au,Mai:2012dt,Guo:2012vv,Mai:2014xna} are being available with the constraint from the SIDDHARTA data. 
The dimensional regularization is applied to the loop function $G_{i}$ with the finite part being specified by the subtraction constant.
Adjusting the subtraction constant adequately, the experimental data such as scattering cross sections, threshold branching ratios and the scattering length can be reproduced well. 
Although there are other regularization schemes constrained by the crossing symmetry~\cite{Lutz:2001yb} and the SU(3) symmetry~\cite{Lutz:1997wt}, 
the present phenomenological regularization scheme is sufficient to consider the $\bar{K}N$ scattering near the threshold.

The $\bar{K}N$ forward scattering amplitude $F_{\bar{K}N}$ is related to the amplitude $T_{ij}$ as 
\begin{align}
F_{\bar{K}N}(\sqrt{s}) = -\frac{M_N}{4\pi\sqrt{s}}T_{11}(\sqrt{s}),  \label{eq:FkN}
\end{align}
where $M_N$ represents the nucleon mass. 
In the isospin $I=0$ channel, there are two resonance poles in the $\Lambda(1405)$ energy region, induced by the attractive interactions of the $\bar{K}N$ channel and the $\pi\Sigma$ channel~\cite{Hyodo:2007jq,Jido:2003cb}. In this paper, we refer to the higher (lower) energy pole near the $\bar{K}N$ ($\pi\Sigma$) threshold as $\bar{K}N$ pole ($\pi\Sigma$ pole).

\subsection{Equivalent single-channel potential}
Our aim is to construct the $\bar{K}N$ single-channel interaction for the application to few-body calculations as well as the $\Lambda(1405)$ analysis. In this work, we construct a single-channel local potential which produces the amplitude equivalent to the chiral coupled-channel approach. The coordinate space wave function calculated by the potential is useful to study the spatial structure of $\Lambda(1405)$. In addition, the local potential is easily implemented in the variational calculations of the few-body systems~\cite{Hiyama:2003cu}. 

To this end, we first extract the single-channel $\bar{K}N$ interaction from the coupled-channel scattering equation~\eqref{eq:BetheSal}. We define the effective interaction $V^{\rm eff}_{11}$ as
\begin{align}
V^{\rm eff}_{11} &= \sum_{m=2}^N V_{1m}G_mV_{m1}+\sum_{m,l=2}^N V_{1m}G_mT_{ml}^{(N-1)}G_lV_{l1},  \label{eq:Feshbach} \\
T_{ml}^{(N-1)} &= V_{ml}^{(N-1)}+\sum_{k=2}^N V_{mk}^{(N-1)}G_k^{(N-1)}T_{kl}^{(N-1)}  \notag \\
&= \left[ (V^{(N-1)})^{-1}-G^{(N-1)} \right]^{-1},\ \ m,l=2,3,...,N.  \notag
\end{align}
The quantities with the superscript ($N-1$) are the $(N-1)\times(N-1)$ matrices. Using this single-channel scattering equation $T_{11}=\left[ (V^{\rm eff}_{11})^{-1}-G_1 \right]^{-1}$, the original amplitude is exactly reproduced. Because of the elimination of the lower energy $\pi\Sigma$ channel, the effective $\bar{K}N$ interaction $V^{\rm eff}_{11}$ has an imaginary part.

Next, we define the energy dependent local potential
\begin{align}
U(r,E) &= g(r)N(E)V^{\rm eff}_{11}(E+M_N+m_K),  \label{eq:equivpot0}  \\
N(E) &= \frac{M_N}{2(E+M_N+m_K)}\frac{\omega_K+E_N}{\omega_K E_N},
\end{align}
where $E$, $E_N$ and $\omega_K$ are the nonrelativistic energy, the energy of the nucleon and the energy of the anti-kaon,
\begin{align}
E &= \sqrt{s}-M_N-m_K,  \notag \\
E_N &= \frac{s-m_K^2+M_N^2}{2\sqrt{s}},  \notag \\
\omega_K &= \frac{s-M_N^2+m_K^2}{2\sqrt{s}},  \notag
\end{align}
with the mass of the antikaon $m_K$. The spatial distribution of the potential is governed by $g(r)$ which is normalized as $\int d{\bm r}\ g(r)=1$. The flux factor $N(E)$ is determined by the matching with the original amplitude at the $\bar{K}N$ threshold in the Born approximation~\cite{Hyodo:2007jq}. 
In this work, we choose a Gaussian for $g(r)$
\begin{align}
g(r) = \frac{1}{\pi^{3/2}b^3}e^{-r^2/b^2},  \notag
\end{align}
where the parameter $b$ determines the range of the potential.
Using the local potential, we can calculate the wave function from the Schr\"odinger equation,
\begin{align}
-\frac{1}{2\mu}\frac{d^2u(r)}{dr^2}+&U(r,E)u(r) = Eu(r)  \label{eq:schro},  
\end{align}
where $\mu=M_Nm_K/(M_N+m_K)$ is the reduced mass and $u(r)$ is the $s$-wave part of the two-body radial wave function. From the behavior of the wave function at $r\to\infty$, the scattering amplitude $F_{\bar{K}N}$ can be obtained.
In Ref.~\cite{Hyodo:2007jq}, the parameter $b$ was determined to match the amplitude $F_{\bar{K}N}$ with the original amplitude in the $\Lambda(1405)$ resonance region. In this work, we determine the parameter $b$ by the matching of the full amplitude at the $\bar{K}N$ threshold. This prescription is along the same line with the determination of the flux factor $N(E)$.

The potential \eqref{eq:equivpot0} well reproduces the original amplitude near the $\bar{K}N$ threshold, while the deviation increases in the energy region far below the threshold. To enlarge the applicability of the potential, we add the correction $\Delta V(E)$ to the strength of the potential,
\begin{align}
U(r,E) &= g(r)N(E)\left[V^{\rm eff}_{11}(E+M_N+m_K)+\Delta V(E) \right].  \label{eq:equivpot1}
\end{align}
For the analytic continuation of the amplitude in the complex energy plane, it is useful to parameterize the strength of the potential by a polynomial in the energy,
\begin{align} 
U(r=0,E) = g(r=0)N(E)\left[ \sum_i K_i \left( \frac{E}{100\ {\rm MeV}} \right)^i \right].  \label{eq:fit} 
\end{align}
We refer to the energy range where the potential is parameterized as parameterized range, which will be specified for each potential. 
We comment on the analytic behavior of the amplitude calculated from the potential~\eqref{eq:fit}. Because the potential is constructed to reproduce the original amplitude, the correct analytic behavior is guaranteed within the parametrized range on the real axis. On the other hand, the extrapolation of this potential to other energy regions should be carefully performed, since some unphysical singularities can in general be developed. This will be discussed in detail in the next section.

\section{Potential construction}  \label{sec:precisepot}  

In this section, we study how the original amplitude is reproduced by the $\bar{K}N$ local potential. 
Examining the previous method in Ref.~\cite{Hyodo:2007jq} in detail, we improve the construction procedure to reproduce the original amplitude even in the complex energy plane. Here, we mainly employ the amplitude of the HNJH model~\cite{Hyodo:2002pk,Hyodo:2003qa} for the comparison with Ref.~\cite{Hyodo:2007jq}. Inclusion of the SIDDHARTA constraint will be discussed in the next section to construct a realistic $\bar{K}N$ potential.

\subsection{Precision of potential in the complex plane}

A resonance state is represented by a pole of the scattering amplitude in the complex energy plane. The pole structure of the $\bar{K}N$ amplitude is therefore important for the study of the spacial structure of $\Lambda(1405)$. It is considered that the pole structure of the $\bar{K}N$ system may affect the result of the $\bar{K}NN$ system \cite{Ikeda:2010tk}. We thus focus on the scattering amplitude from the previous potential in the complex plane. 

\begin{figure}[tb]
\begin{center}
\includegraphics[width=8cm,bb=0 0 985 714]{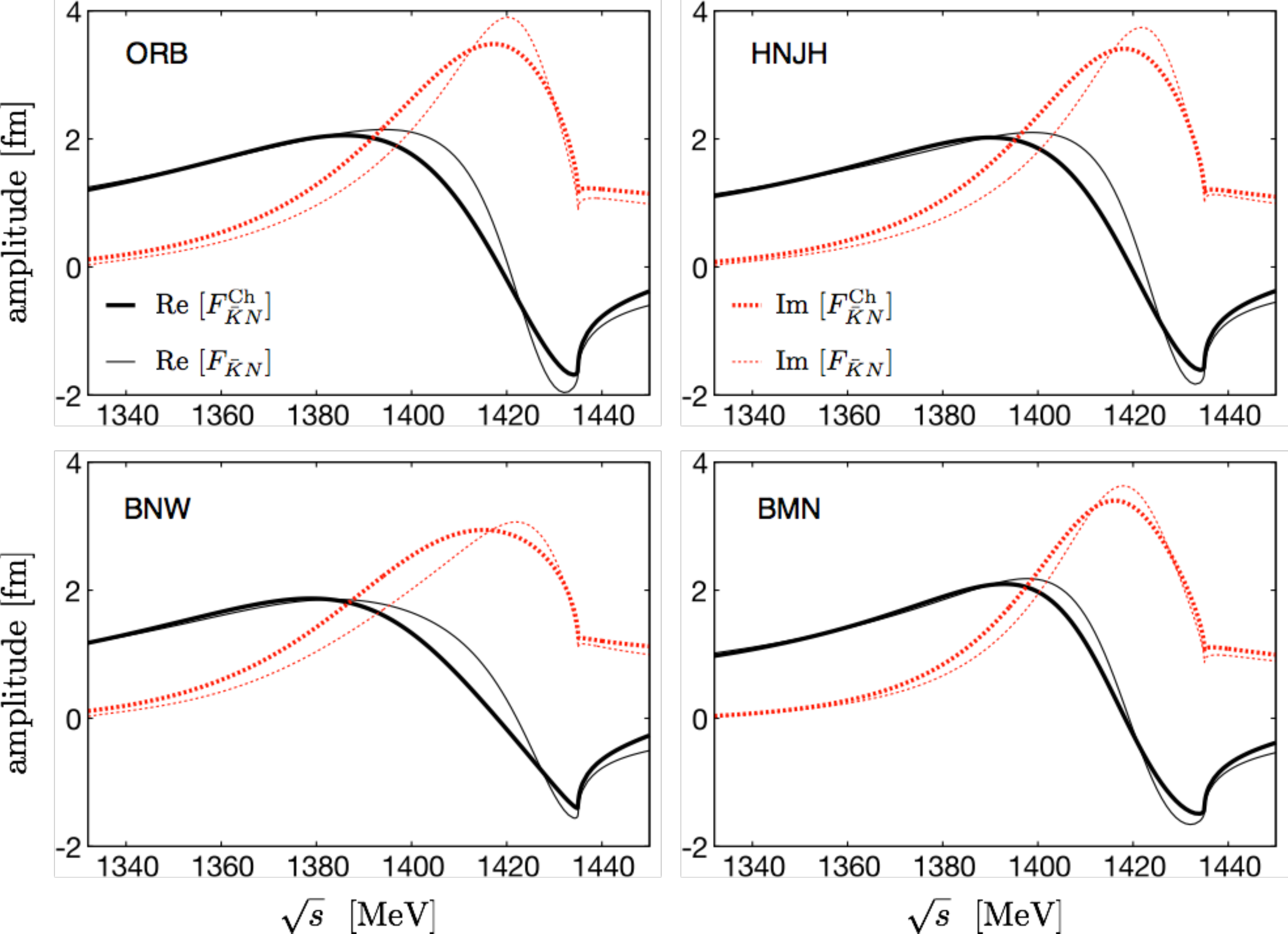}
\caption{(Color online) Scattering amplitudes from the local potentials $F_{\bar{K}N}$ (thick lines) and the amplitudes directly from chiral unitary approach $F_{\bar{K}N}^{\rm Ch}$ (thin lines) with models ORB~\cite{Oset:2001cn}, HNJH~\cite{Hyodo:2002pk,Hyodo:2003qa}, BNW~\cite{Borasoy:2005fq,Borasoy:2005ie} and BMN~\cite{Borasoy:2006sr}. The real (imaginary) parts are shown by the solid (dotted) lines.}
\label{fig:Vequiv}  
\end{center}
\end{figure}

In Fig.~\ref{fig:Vequiv}, we compare the $\bar{K}N\ (I=0)$ scattering amplitude from the local potential $F_{\bar{K}N}$ in Ref.~\cite{Hyodo:2007jq} with the corresponding original amplitude in the chiral unitary approach $F^{\rm Ch}_{\bar{K}N}$ for the models ORB~\cite{Oset:2001cn}, HNJH~\cite{Hyodo:2002pk,Hyodo:2003qa}, BNW~\cite{Borasoy:2005fq,Borasoy:2005ie} and BMN~\cite{Borasoy:2006sr} on the real axis.
The $\bar{K}N$ amplitudes on the real axis are reasonably well reproduced by the potentials in Ref.~\cite{Hyodo:2007jq}. On the other hand, we find a large deviation of the amplitude in the complex energy plane. In Table~\ref{tab:pole_FkN_FkNCh}, we list the pole positions of the scattering amplitudes. While chiral unitary approaches generate two poles in the $\Lambda(1405)$ energy region, the local potentials give only one pole. In addition, the position of the pole does not agree with neither of the original poles. Hence, the potential construction procedure should be improved by paying attention to the amplitude in the complex energy plane.

%
\begin{table}[tb]
\begin{center}
\begin{ruledtabular}
\begin{tabular}{lll}
Model  &  \multicolumn{2}{c}{pole position [MeV]}  \\ \cline{2-3}
  &  $F_{\bar{K}N}^{\rm Ch}$  &  $F_{\bar{K}N}$  \\ \hline
ORB \cite{Oset:2001cn}  &  $1427-17i$, $1389-64i$  &  $1419-42i$  \\
HNJH \cite{Hyodo:2002pk,Hyodo:2003qa}  &  $1428-17i$, $1400-76i$  &  $1421-35i$  \\
BNW \cite{Borasoy:2004kk,Borasoy:2005ie}  &  $1434-18i$, $1388-49i$  &  $1404-46i$  \\
BMN \cite{Borasoy:2006sr}  &  $1421-20i$, $1440-76i$
  &  $1416-27i$  \\
\end{tabular}
\caption{Pole positions of the original scattering amplitudes from chiral unitary approach $F_{\bar{K}N}^{\rm Ch}$ and the amplitudes from the local potentials $F_{\bar{K}N}$. All poles are found in the $\pi\Sigma$ unphysical and $\bar{K}N$ physical Riemann sheet. The pole at $1440-76i$ in the BMN model is above the $\bar{K}N$ threshold, and hence is not in the most adjacent sheet to the real axis.}
\label{tab:pole_FkN_FkNCh}  
\end{ruledtabular}
\end{center}
\end{table}

To improve the construction procedure, we introduce several quantities to assess the deviation of the amplitudes in the complex plane. For the discussion of $\Lambda(1405)$, we consider that the following energy region is relevant\footnote{The lower boundary of Re[$z$] (1332 MeV) is set at the $\pi\Sigma$ threshold.},
\begin{align}
1332\ {\rm MeV}&\leq {\rm Re}[z] \leq1450\ {\rm MeV}  \notag \\
-100\ {\rm MeV}&\leq {\rm Im}[z] \leq50\ {\rm MeV} , \label{eq:Erange}
\end{align}
where $z$ represents the complex energy of the two-body system. 
First, we define the average deviation $\Delta F_{\rm real}$ between the amplitude from the local potential $F_{\bar{K}N}$ and the amplitude from chiral unitary approach $F_{\bar{K}N}^{\rm Ch}$ on the real energy axis as
\begin{align}
\Delta F_{\rm real} = \frac{\displaystyle\int d\sqrt{s}\ |F_{\bar{K}N}^{\rm Ch}(\sqrt{s})-F_{\bar{K}N}(\sqrt{s})|}{\displaystyle\int d\sqrt{s}\ |F_{\bar{K}N}^{\rm Ch}(\sqrt{s})|} .  \label{eq:deltaFreal}
\end{align} 
When $\Delta F_{\rm real}$ is small, the amplitude on the real axis is well reproduced by the potential. 
When $\Delta F_{\rm real} \sim 1$, it means the average deviation on the real energy axis reaches the same amount as the average magnitude of $|F_{\bar{K}N}|$. With the HNJH model, we obtain $\Delta F_{\rm real}=0.14$.

Second, we define the deviation of the amplitude at complex energy $z$, 
\begin{align}
\Delta F(z) = \left| \frac{F_{\bar{K}N}^{\rm Ch}(z)-F_{\bar{K}N}(z)}{F_{\bar{K}N}^{\rm Ch}(z)} \right| . \label{eq:deltaF}
\end{align}
In this paper, we regard that the amplitude is well reproduced when the deviation is smaller than 20 \%:
\begin{align}
\Delta F(z) < 0.2.  \label{eq:preciserange}
\end{align}
We call the energy region satisfying this condition the ``precise region''. We also define the percentage of this precise region in the relevant energy region~\eqref{eq:Erange} by
\begin{align}
P_{\rm comp} = \frac{\displaystyle\iint d({\rm Re}z) d({\rm Im}z) \ \Theta(0.2-\Delta F(z))}{\displaystyle\iint d({\rm Re}z) d({\rm Im}z)}\times 100.  \label{eq:Pcomp}  
\end{align}
If the local potential well reproduces the original amplitude well in the relevant region of the complex energy plane, then we have $P_{\rm comp}\sim 100$. The HNJH model gives $P_{\rm comp}=50$, which quantifies the insufficiency of the precision in the complex energy plane.

\subsection{Region near the real axis}

\begin{figure}[tb]
\begin{center}
\includegraphics[width=8cm,bb=0 0 603 489]{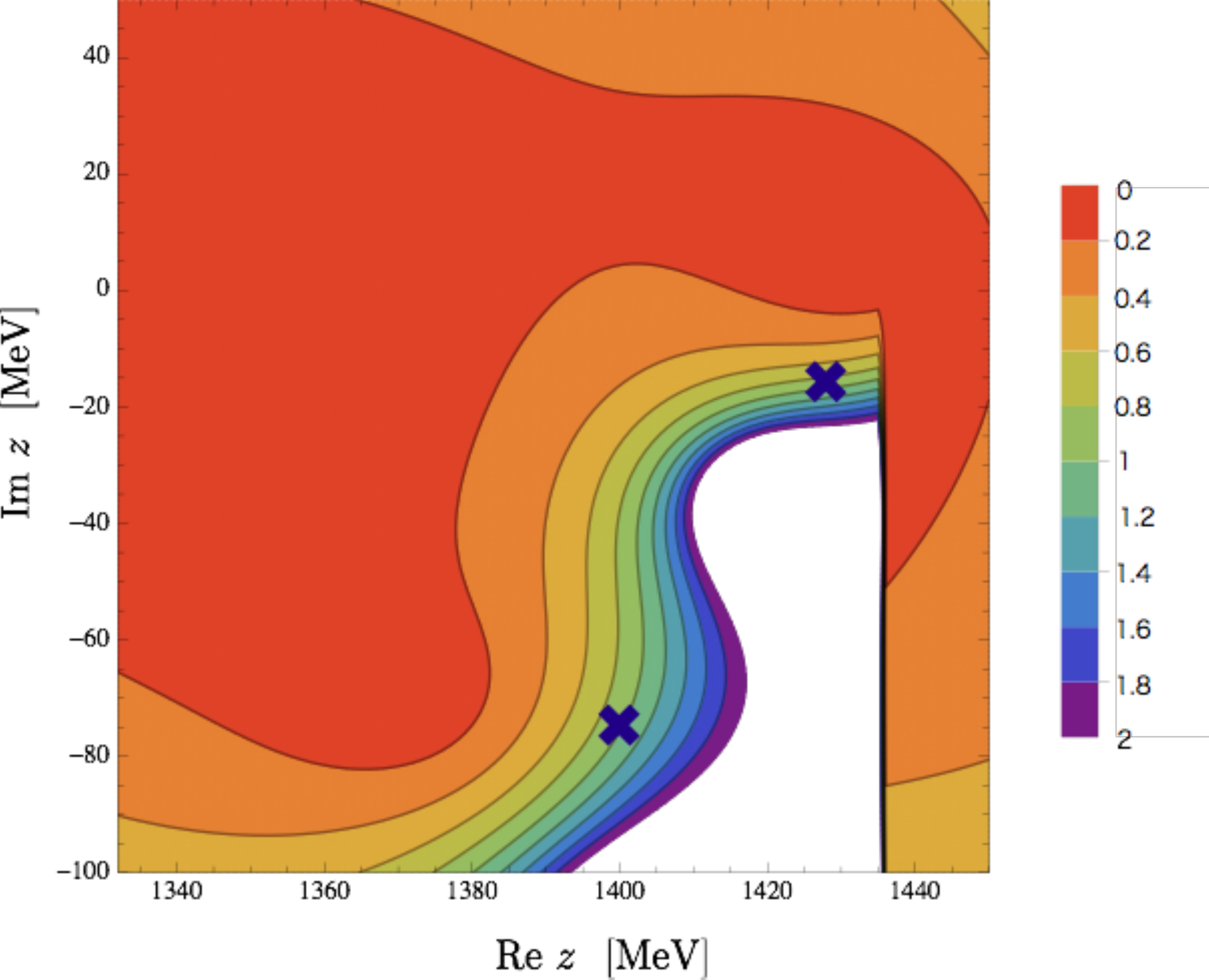}
\caption{(Color online) The contour plot of $\Delta F$ of the HNJH potential in Ref.~\cite{Hyodo:2007jq}. The unfilled region corresponds to large deviation, $\Delta F>2$. Precise region is defined as $\Delta F<0.2$. The crosses represent the original pole positions of $\Lambda(1405)$. }
\label{fig:HWprec}  
\end{center}
\end{figure}

We explain how to reproduce the amplitude in the complex energy plane. We first focus on the region near the real energy axis including the $\bar{K}N$ pole. Here we use the HNJH model as an example%
\footnote{In this work, the range parameter $b$ of the potential is determined as 0.46 fm by the new prescription explained in Sec. \ref{sec:formulation}, in contrast to $b=$0.47 fm of the potential in Ref.~\cite{Hyodo:2007jq}}. 
Let us show the contour plot of $\Delta F$ in the complex energy plane with the potential in Ref.~\cite{Hyodo:2007jq} in Fig.~\ref{fig:HWprec}. 
Here we choose the most adjacent Riemann sheet to the real energy axis.
It is seen from Fig. \ref{fig:HWprec} that the deviation in the region around $\re [z]\sim 1400\ {\rm MeV}$ is larger than the other region. The deviation of the amplitude should influence the pole positions of $\Lambda(1405)$. The reason for the deviation is that the correction to the potential $\Delta V$ has been applied only in the region below $1400\ {\rm MeV}$ in the previous work. Furthermore, the $\Delta V$ has been chosen to be real, based on the dominance of the real part in $V^{\rm eff}_{11}$.

In this work, we add $\Delta V$ in the relevant energy region for the $\bar{K}N$ pole, 1332-1450 MeV. Hereafter we call the region where $\Delta V$ is applied the correction range. To reproduce the original amplitude near the $\Lambda(1405)$ resonance region,we introduce the complex correction $\Delta V$. As a consequence, $\Delta F_{\rm real}$ is significantly reduced. We call the new potential with the complex $\Delta V$ ``Potential I'' and summarize its properties in Table~\ref{tab:improvedpot} 
together with the property of the corresponding potential in Ref.~\cite{Hyodo:2007jq}.

\begin{figure}[tb]
\begin{center}
\includegraphics[width=8cm,bb=0 0 603 489]{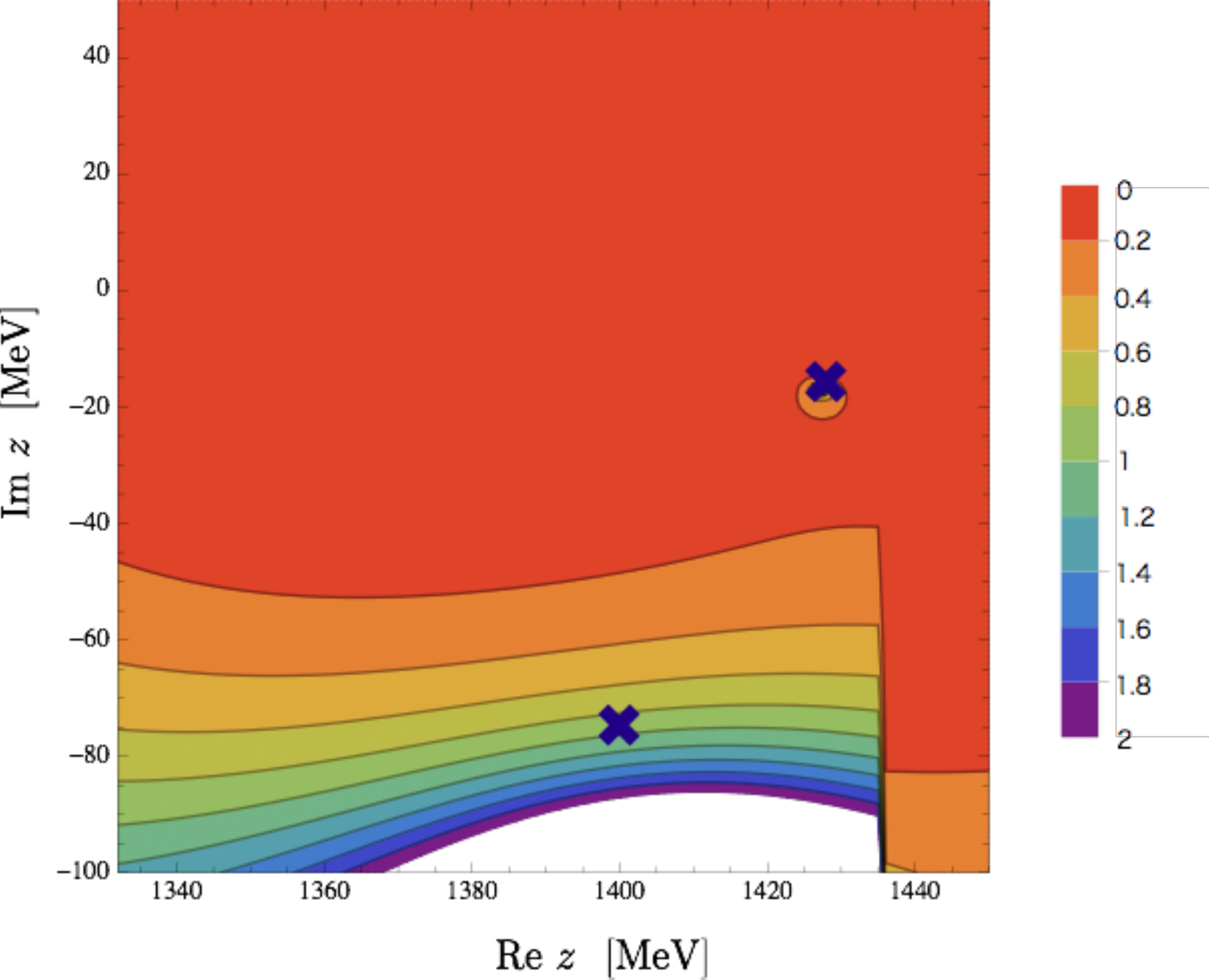}
\caption{(Color online) The contour plot of $\Delta F$ of Potential I. The unfilled region corresponds to large deviation, $\Delta F>2$. Precise region is defined as $\Delta F<0.2$. The crosses represent the original pole positions of $\Lambda(1405)$. }
\label{fig:pot1prec}  
\end{center}
\end{figure}
%
\begin{table*}[tb]
\begin{center}
\begin{ruledtabular}
\begin{tabular}{cccc}
&  Ref.~\cite{Hyodo:2007jq}  &  Potential I  &  Potential II  \\ \hline
$b$ [fm]  &  0.47  &  0.46  &  0.46  \\ 
$\Delta V$  &  real  & complex  &  complex  \\ 
polynomial type  &  third order &  third order &   tenth order  \\
correction range  &  1300-1400  &  1332-1450  &  1332-1521  \\ 
parameterized range  &  1300-1450  &  1332-1450  &  1332-1521  \\
$\Delta F_{\rm real}$   &  $1.4\times10^{-1}$  &  $4.8\times10^{-3}$  &  $4.0\times10^{-4}$  \\ 
$P_{\rm comp}$   &  50  &  68  &  85  \\ 
pole positions [MeV]  &  $1421-35i$  &  $1427-17i$  &  $1428-17i,\ 1400-77i$  \\
\end{tabular} 
\caption{Properties of the HNJH potential in Ref.~\cite{Hyodo:2007jq} and  Potential I and Potential II in this work. Shown are the potential range parameters $b$, the corrections to the strength of the potentials $\Delta V$, the polynomial types of the potential strength in energy, the correction ranges where $\Delta V$ is applied, the parameterized ranges by the polynomials, the average deviations $\Delta F_{\rm real}$ from the amplitudes of chiral unitary approach $F_{\bar{K}N}^{\rm Ch}$ on the real energy axis, the percentages of the precise region in the complex energy plane, and the pole positions of the amplitudes from the potentials $F_{\bar{K}N}$. The pole positions of the original amplitude $F_{\bar{K}N}^{\rm Ch}$ are $1428-17i$ MeV and $1400-76i$ MeV.}
\label{tab:improvedpot}  
\end{ruledtabular}
\end{center}
\end{table*}

With Potential I, the deviation on the real energy axis $\Delta F_{\rm real}$ is reduced by two orders of magnitude. Thanks to the reduction of $\Delta F_{\rm real}$, the $\bar{K}N$ pole position is also significantly improved. We show the contour plot of $\Delta F$ in Fig.\ref{fig:pot1prec}. Comparing Fig.~\ref{fig:HWprec} with Fig.~\ref{fig:pot1prec}, we find that the precise region ($\Delta F<0.2$) of Potential I satisfying Eq.~\eqref{eq:preciserange} is extended over the $\bar{K}N$ pole. The improvement of the pole position can be understood by this enlargement of the precise region.
Quantitatively, $P_{\rm comp}$ in Eq.~\eqref{eq:Pcomp} increases from 50 to 68. In this way, the precision near the real axis can be improved by introducing the complex correction $\Delta V$ in the relevant correction range.

\subsection{Region far from the real axis}
While Potential I reproduces the original amplitude near the real energy axis, the deviation of the amplitude increases in the region far from the real axis (see Fig.~\ref{fig:pot1prec}) and the $\pi\Sigma$ pole does not appear. Here we further improve the potential paying attention to the region far from the real axis. 

In principle, if the original amplitude is completely reproduced in the whole range on the real energy axis, the analytic continuation in the complex energy plane is unique. This suggests that the increase of the parameterized range will improve the precision of the potential far from the real axis.\footnote{In this subsection, the correction range is chosen to be the same with the parameterized range.} On the other hand, there is a limitation of extension of the parameterized range because of the threshold effect. In the present framework of the effective single-channel potential with polynomial parameterization, it is difficult to incorporate the non-analytic threshold effect of the other channels. The parameterized range can only be extended to the nearest thresholds. In this case, the parameterization of the $\bar{K}N$ potential strength should be performed between the $\pi\Sigma$ threshold (1331 MeV) and the $\eta\Lambda$ threshold (1664 MeV). In order to keep the precision on the real axis for the larger parameterized range, we increase the degree of the polynomial from the third order to the tenth order. 

\begin{figure}[tb]
\begin{center}
\includegraphics[width=8cm,bb=0 0 603 489]{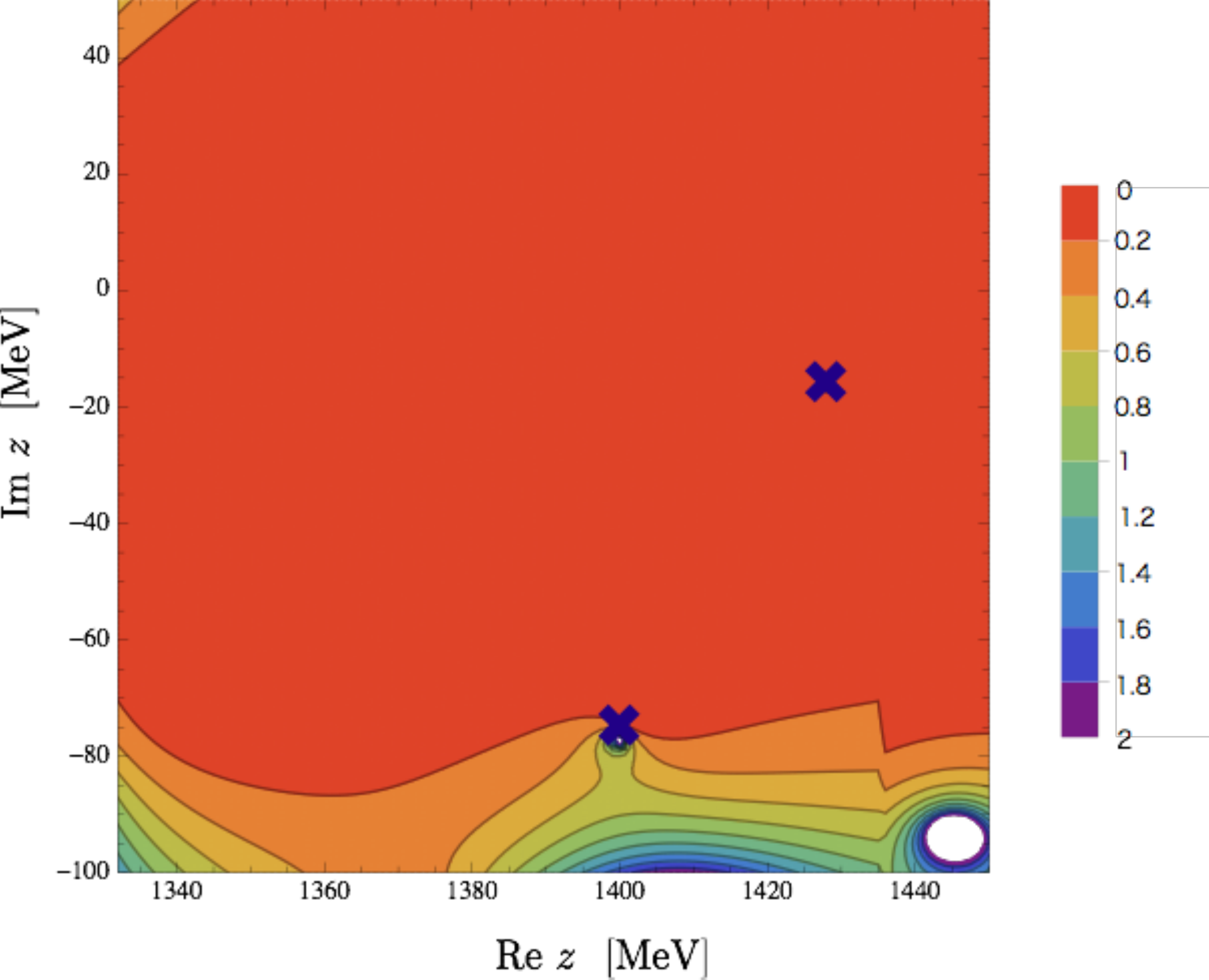}
\caption{(Color online) The contour plot of $\Delta F$ of Potential II. The unfilled region corresponds to large deviation, $\Delta F>2$. Precise region is defined as $\Delta F<0.2$. The crosses represent the original pole positions of $\Lambda(1405)$. }
\label{fig:pot2prec}  
\end{center}
\end{figure}

To examine the above strategy, we construct the potentials varying the parameterized range by 1 MeV. The typical results of $\Delta F_{\rm real}$, $P_{\rm comp}$, and the pole positions of these potentials are shown in Table \ref{tab:changefit}.
In all cases, $\Delta F_{\rm real}$ is reduced by an order of magnitude from that of Potential I. This is because we change the parameterization from the third order to the tenth order polynomial. Though the wider fitting range leads to the slightly larger $\Delta F_{\rm real}$, the order of magnitude remains same.
In general, when a high-degree polynomial is used for the parameterization, artificial poles appear between the $\bar{K}N$ and $\pi\Sigma$ thresholds. In the present case, this occurs when the fitting range is smaller than $\sim$ 1500 MeV. However, as the fitting range increases, these unphysical poles move away from the relevant energy region and only two physical poles remain. The $\bar{K}N$ pole appears at the original pole position, $1428-17i$ MeV and is stable against the variation of the parameterized range. On the other hand, the position of the $\pi\Sigma$ pole depends on the parametrized range. The optimized value of the upper boundary of the parameterized range is 1521 MeV to reproduce the original pole position, $1400-76i$ MeV. At the same time, the maximum value of $P_{\rm comp}$ is achieved. We call the potential with the best parametrized range Potential II.
We show the contour plot of $\Delta F$ with Potential II in Fig.\ref{fig:pot2prec}.
As shown in Fig.\ref{fig:pot2prec}, we succeed in extending the precise region to ${\rm Im} z\sim -80\ {\rm MeV}$, near the $\pi\Sigma$ pole. As a consequence, we obtain two poles both at the correct positions. 

\begin{table*}[tb]
\begin{center}
\begin{ruledtabular}
\begin{tabular}{cccl} 
upper boundary [MeV]  &  $\Delta F_{\rm real}$  &  $P_{\rm comp}$ &  pole positions [MeV]  \\ \hline
1450  &  $1.8\times10^{-4}$  &  59  &  $1428-17i,\ 1388-60i$, unphysical poles  \\
1500  &  $2.6\times10^{-4}$  &  71  &  $1428-17i,\ 1404-70i$, unphysical poles  \\
1521  &  $4.0\times10^{-4}$  &  85  &  $1428-17i,\ 1400-77i$  \\
1550  &  $5.9\times10^{-4}$  &  79  &  $1428-17i,\ 1392-82i$  \\
1600  &  $6.8\times10^{-4}$  &  77  &  $1428-17i,\ 1389-83i$  \\
1650  &  $8.8\times10^{-4}$  &  77  &  $1428-17i,\ 1389-87i$  \\    
\end{tabular} 
\caption{Results of the precision of the potentials against the variation of the parameterized range for the HNJH model. Shown are the average deviations $\Delta F_{\rm real}$ from the amplitudes of chiral unitary approach $F_{\bar{K}N}^{\rm Ch}$ on the real energy axis, the percentages of the precise region in the complex energy plane, and the pole positions of the amplitudes from the potentials $F_{\bar{K}N}$. The ``unphysical poles'' stand for the artificial poles generated between the $\bar{K}N$ and $\pi\Sigma$ thresholds as explained in the text. The pole positions of the original amplitude $F_{\bar{K}N}^{\rm Ch}$ are $1428-17i$ MeV and $1400-76i$ MeV.}
\label{tab:changefit}  
\end{ruledtabular}
\end{center}
\end{table*}

It turns out that the largest parameterized range does not always lead to the best potential. In the present case, this is because the $\pi\Sigma$ pole position moves along with the change of the parameterized range. The best potential is achieved when the $\pi\Sigma$ pole comes closest to the original position.

\section{Application}  \label{sec:application}  

In the previous section, we have established the construction procedure to reproduce the original amplitude in the complex energy plane, considering the high precision on the real energy axis and the wider parameterized range. In this section, we apply this procedure to chiral unitary approach with SIDDHARTA constraint~\cite{Ikeda:2011pi,Ikeda:2011pi,Ikeda:2012au} and construct the realistic $\bar{K}N$ local potential. This new potential is then used to estimate the mean distance between $\bar{K}$ and nucleon, that is, the spatial structure of $\Lambda(1405)$. 

\subsection{Realistic $\bar{K}N$ potential}

As we explained in Sec.~\ref{sec:intro}, the constraint from the precise SIDDHARTA data is crucial for the quantitative calculation of the $\bar{K}$ and nucleons systems. In this section, we construct the $\bar{K}N$ local potential based on the amplitude of Refs.~\cite{Ikeda:2011pi,Ikeda:2012au} with the SIDDHARTA constraint. To apply to the few-body $\bar{K}$-nuclei, we construct the potential of the $I=1$ amplitude in addition to the $I=0$ channel.

\begin{figure}[tb]
\begin{center}
\includegraphics[width=8cm,bb=0 0 664 475]{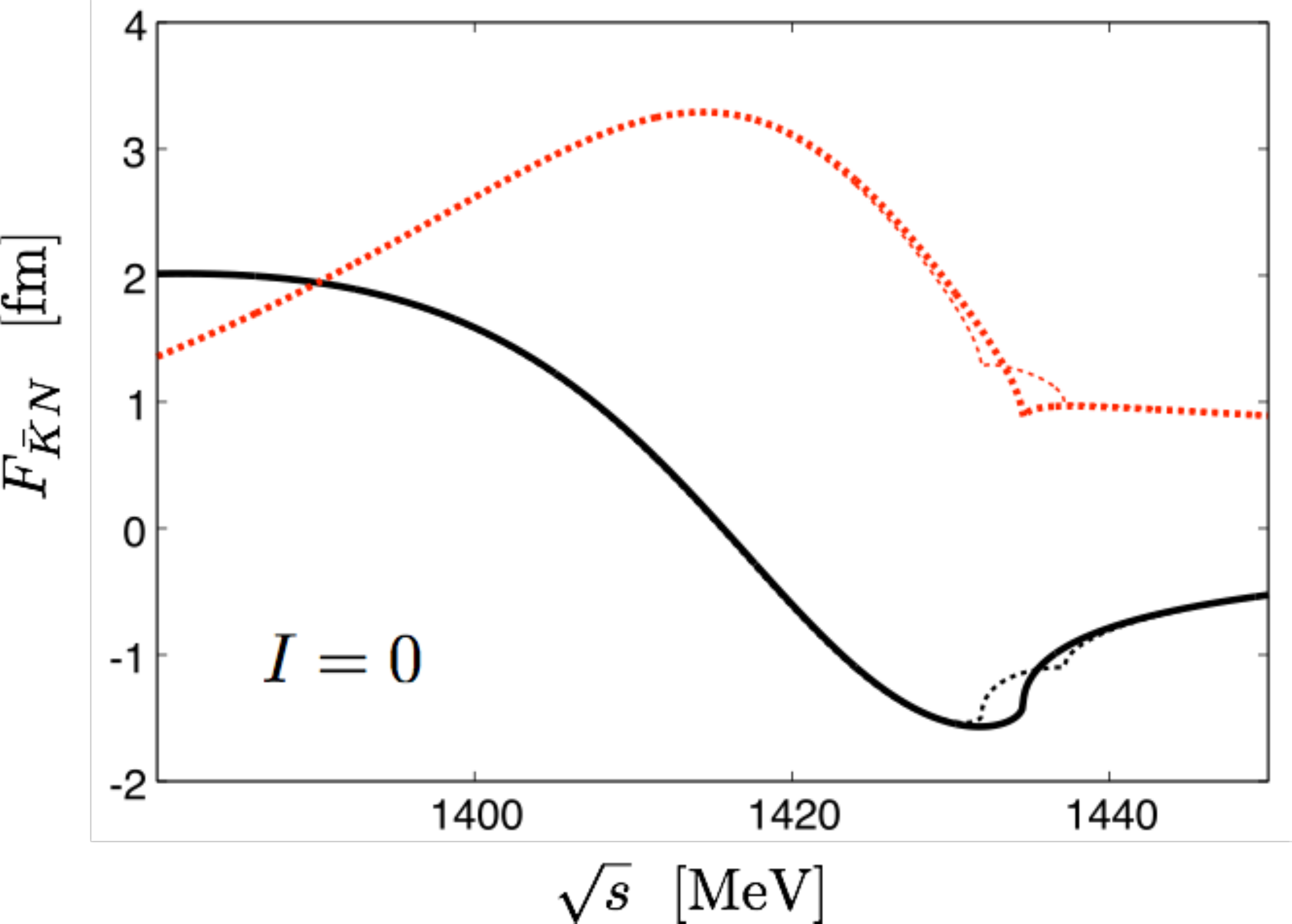}
\caption{(Color online) The result of the isospin symmetric $\bar{K}N$ amplitude in the $I=0$ channel (thick lines) and the combination of the original amplitudes $(F_{K^{-}pK^{-}p}+2F_{K^{-}p\bar{K}^{0}n}+F_{\bar{K}^{0}n\bar{K}^{0}n})/2$ with isospin breaking~\cite{Ikeda:2011pi,Ikeda:2012au} (thin lines). The real parts are shown by the solid lines, and the imaginary parts are shown by the dotted lines.}
\label{fig:isobreaking}  
\end{center}
\end{figure}

The amplitude of Refs.~\cite{Ikeda:2011pi,Ikeda:2012au} is given in the particle basis with the isospin breaking effect in the hadron masses. 
On the other hand, the potential in the isospin basis with isospin symmetry is useful for various applications. Moreover, in the practical potential construction procedure, the existence of multiple thresholds in the particle basis prevents us from enlarging the parameterized range.
We thus construct the isospin symmetric $\bar{K}N$ amplitude by replacing the physical hadron masses by the isospin averaged ones keeping the low energy constants and subtraction constants the same as Refs.~\cite{Ikeda:2011pi,Ikeda:2012au}. The result of the isospin symmetric $\bar{K}N$ amplitude $(I=0)$ is shown in Fig.~\ref{fig:isobreaking} together with the combination of the original amplitudes $(F_{K^{-}pK^{-}p}+2F_{K^{-}p\bar{K}^{0}n}+F_{\bar{K}^{0}n\bar{K}^{0}n})/2$ of Refs.~\cite{Ikeda:2011pi,Ikeda:2012au}. The difference stems from the isospin breaking effect.
From this figure, we find that these amplitudes well agree with each other except for the tiny region near the $\bar{K}N$ threshold. 
Since the difference in the most important region for $\Lambda(1405)$ and the $\bar{K}NN$ systems is negligible, we adopt this isospin symmetric amplitude to construct the $\bar{K}N$ potential. 

\begin{figure}[tb]
\begin{center}
\includegraphics[width=8cm,bb=0 0 603 489]{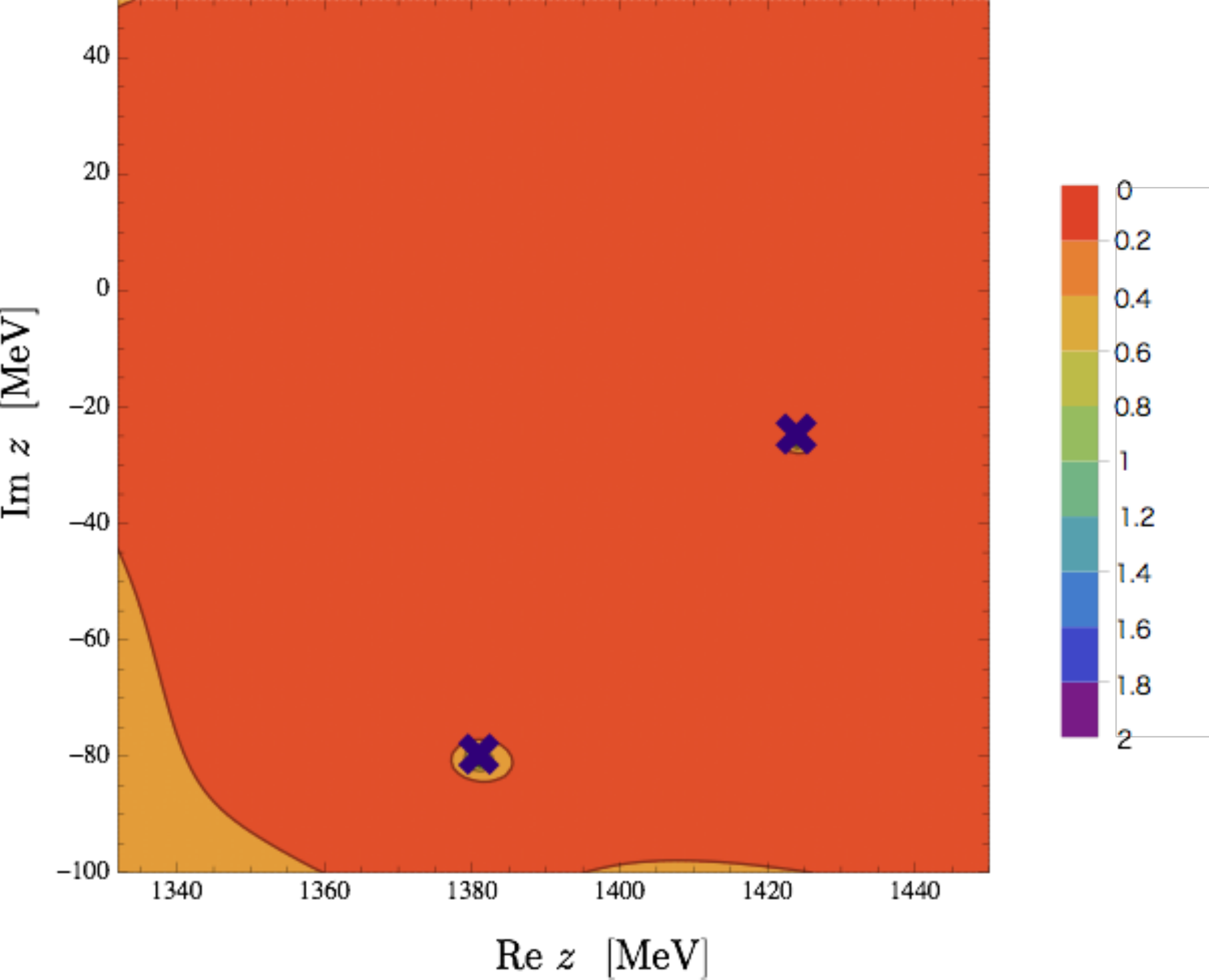}
\caption{(Color online) The contour plot of $\Delta F$ of SIDDHARTA potential $(I=0)$. Precise region is defined as $\Delta F<0.2$. The crosses represent the original pole positions of $\Lambda(1405)$. }
\label{fig:sidprec}  
\end{center}
\end{figure}

\begin{figure}[tb]
\begin{center}
\includegraphics[width=8cm,bb=0 0 736 529]{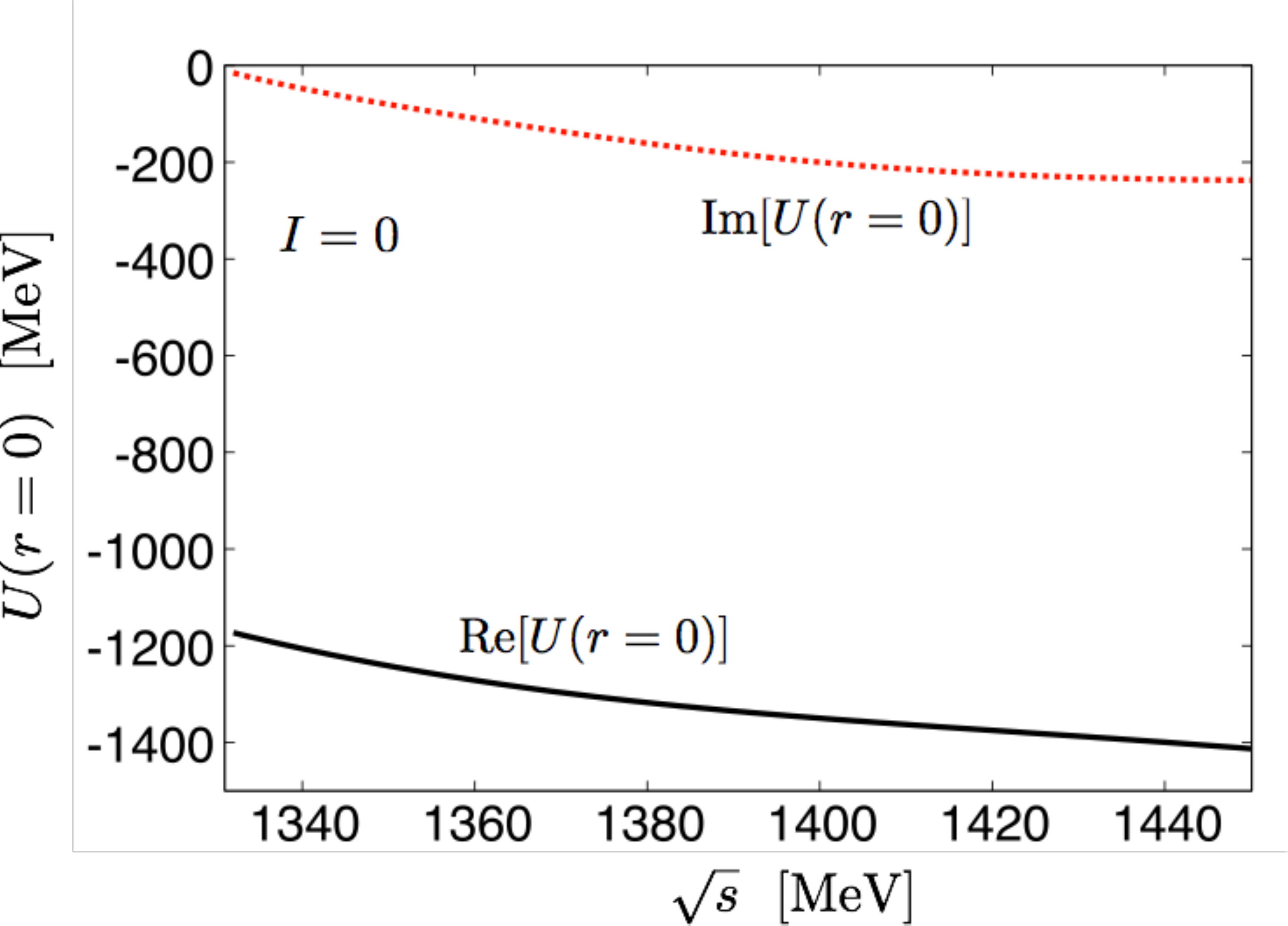}
\caption{(Color online) Strength of SIDDHARTA potential $(I=0)$ $U(r,E)$ at $r=0$. The real part is shown by the solid line, and the imaginary part is shown by the dotted line.}
\label{fig:sidpoteI=0}  
\end{center}
\end{figure}

Following the construction procedure in Sec.~\ref{sec:formulation}, here we determine the Gaussian parameter $b=0.38\ {\rm fm}$. We show the properties of the potentials with various parameterized ranges in Table~\ref{tab:sidfit}. The optimal upper boundary of the parameterized range to reproduce the pole positions is found to be 1657 MeV. We call the best potential SIDDHARTA potential $(I=0)$. The properties of the SIDDHARTA potential $(I=0)$ are summarized in Table~\ref{tab:sidI=0} and the contour plot of $\Delta F$ is shown in Fig.~\ref{fig:sidprec}.
We find that SIDDHARTA potential $(I=0)$ well reproduces the original amplitude in the complex energy plane ($\Delta F_{\rm real}=5.4\times 10^{-3}$, $P_{\rm comp}=96$), and the poles of $\Lambda(1405)$ appear at the same position in the accuracy of 1 MeV.\footnote{We note that the maximum of $P_{\rm comp}$ is achieved when the upper boundary is set to be 1658 MeV. Because the value of $P_{\rm comp}$ depend on the definition of the relevant region~\eqref{eq:Erange}, we determine the best potential by the accuracy of the pole positions.} The strength of the potential is shown in Fig.~\ref{fig:sidpoteI=0} as a function of the energy. The energy dependence of the potential strength is not strong, but is important to precisely reproduce the original amplitude. The coefficients of the strength $K_i$ in Eq.~\eqref{eq:fit} are shown in Table~\ref{tab:KisidI0}.

\begin{table*}[bt]
\begin{center}
\begin{ruledtabular}
\begin{tabular}{cccl} 
upper boundary\ [MeV] & $\Delta F_{\rm real}$ & $P_{\rm comp}$ & pole positions\ [MeV] \\ \hline
1450  &  0.91$\times10^{-3}$ &  50  &  $1424-28i,\ 1381-49i$, unphysical poles  \\ 
1500  &  1.7$\times10^{-3}$   &  62  &  $1424-26i,\ 1395-62i$, unphysical poles  \\
1550  &  2.4$\times10^{-3}$   &  70  &  $1424-26i,\ 1379-68i$  \\
1600  &  2.8$\times10^{-3}$   &  72  &  $1424-26i,\ 1381-70i$  \\
1650  &  4.8$\times10^{-3}$   &  86  &  $1424-26i,\ 1382-79i$  \\ 
1657  &  5.4$\times10^{-3}$   &  96  &  $1424-26i,\ 1381-81i$  \\    
\end{tabular}  
\caption{Results of the precision of the potentials against the variation of the parameterized range for the amplitude with SIDDHARTA constraint. Shown are the average deviations $\Delta F_{\rm real}$ from the amplitudes of chiral unitary approach $F_{\bar{K}N}^{\rm Ch}$ on the real energy axis, the percentages of the precise region in the complex energy plane, and the pole positions of the amplitudes from the potentials $F_{\bar{K}N}$. The ``unphysical poles'' stand for the artificial poles generated between the $\bar{K}N$ and $\pi\Sigma$ thresholds as explained in the text. The pole positions of the original amplitude $F_{\bar{K}N}^{\rm Ch}$ are $1424-26i$ MeV and $1381-81i$ MeV.}
\label{tab:sidfit} 
\end{ruledtabular}
\end{center}
\end{table*}
%
\begin{table}[bt]
\begin{center}
\begin{tabular}{cc} \hline\hline
 & SIDDHARTA potential $(I=0)$  \\ \hline
$b$\ [fm]  &  0.38    \\ 
$\Delta V$  &  complex     \\ 
polynomial type &  tenth order  \\ 
parameterized range\ [MeV]  &  1332-1657  \\ 
$\Delta F_{{\rm real}}$\  & $5.4\times 10^{-3}$  \\ 
$P_{{\rm comp}}$\   &  96   \\
pole positions\ [MeV]  &  $1424-26i,\ 1381-81i$    \\ \hline\hline
\end{tabular} 
\caption{Properties of SIDDHARTA potential $(I=0)$. Shown are the potential range parameters $b$, the corrections to the strength of the potential $\Delta V$, the polynomial type of the potential strength in energy, the parameterized range by the polynomials, the average deviation $\Delta F_{\rm real}$ from the amplitude of chiral unitary approach $F_{\bar{K}N}^{\rm Ch}$ on the real energy axis, the percentage of the precise region in the complex energy plane, and the pole positions of the amplitude from the potential $F_{\bar{K}N}$. The pole positions of the original amplitude $F_{\bar{K}N}^{\rm Ch}$ are $1424-26i$ MeV and $1381-81i$ MeV.}
\label{tab:sidI=0} 
\end{center}
\end{table}
%

\begin{table*}[placement specifier]
\begin{center}
{\tabcolsep = 2.7mm
\begin{tabular}{cccccccccc} \hline\hline
\multicolumn{2}{c}{$K_0$ [fm]}  &  \multicolumn{2}{c}{$K_1$ [fm]}  &  \multicolumn{2}{c}{$K_2$ [fm]}  &  \multicolumn{2}{c}{$K_3$ [fm]} \\ 
Re  &  Im  &  Re  &  Im  &  Re  &  Im  &  Re  &  Im   \\ 
$-10.833$ & $-1.8149$ & $-2.7962$ & $-0.64315$ & $-0.47980$ & $0.88991$ & $-0.64480$ & $-0.55225$  \\ \hline

\multicolumn{2}{c}{$K_4$ [fm]}   &  \multicolumn{2}{c}{$K_5$ [fm]}  &  \multicolumn{2}{c}{$K_6$ [fm]}  &  \multicolumn{2}{c}{$K_7$ [fm]}   \\ 
Re  &  Im  &  Re  &  Im  &  Re  &  Im  &  Re  &  Im   \\
 $0.44645$ & $0.0048399$ & $0.089658$ & $0.47326$ & $-0.23222$ & $-0.38284$ & $0.027650$ & $-0.072843$   \\ \hline
 
 \multicolumn{2}{c}{$K_8$ [fm]}  &  \multicolumn{2}{c}{$K_9$ [fm]}  &  \multicolumn{2}{c}{$K_{10}$ [fm]}  &  \multicolumn{2}{c}{}  \\ 
Re  &  Im  &  Re  &  Im  &  Re  &  Im  &  \multicolumn{2}{c}{}  \\ 
$0.059123$ & $0.22152$ & $-0.024071$ & $-0.099375$ & $0.0022208$ & $0.014415$   &  \multicolumn{2}{c}{}  \\ \hline\hline
\end{tabular}  }
\caption{Coefficients $K_i$ in Eq.~\eqref{eq:fit} of the strength of SIDDHARTA potential $(I=0)$.}
\label{tab:KisidI0} 
\end{center}
\end{table*}
%

\begin{figure}[tb]
\begin{center}
\includegraphics[width=8cm,bb=0 0 724 532]{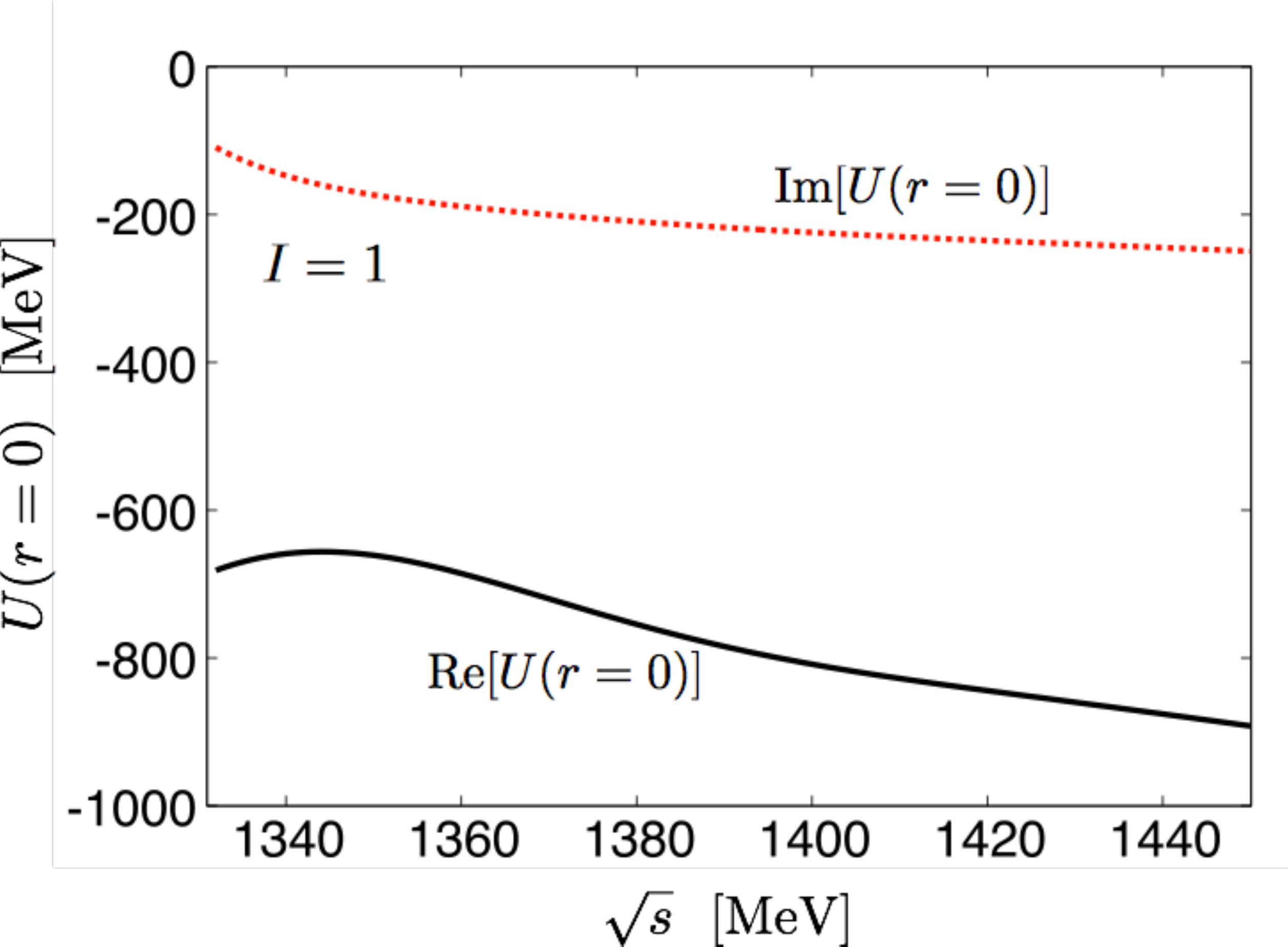}
\caption{(Color online) Strength of SIDDHARTA potential $(I=1)$ $U(r,E)$ at $r=0$. The real part is shown by the solid line, and the imaginary part is shown by the dotted line.}
\label{fig:sidpoteI=1}  
\end{center}
\end{figure}

In the same way, we construct the $\bar{K}N$ local potential for the $I=1$ channel from the combination of $(F_{K^{-}pK^{-}p}-2F_{K^{-}p\bar{K}^{0}n}+F_{\bar{K}^{0}n\bar{K}^{0}n})/2$. The range parameter of the potential is determined to be $b=0.37$ fm.
In this channel, however, the natural analytic continuation of the amplitude is not possible, because of the prescription to avoid the unphysical cut of the amplitude \cite{Borasoy:2005ie}. In contrast to the $I=0$ channel, the best value of the upper bound of the parameterized range cannot be determined from the information of the complex energy plane. Here we use the same parameterized range as that in the $I=0$ channel. This may be sufficient for the present purpose because the interaction in this channel is not as strong as the $I=0$ channel and the contribution to few-body systems is considered to be small. 
In this way, we construct SIDDHARTA potential $(I=1)$ whose strength at $r=0$ and the coefficients $K_i$ are shown in Fig.~\ref{fig:sidpoteI=1} and Table~\ref{tab:KisidI1}, respectively. As expected, the strength of the real part of the potential is smaller than the $I=0$ counterpart. The imaginary part is similar magnitude with $I=0$, suggesting the absorption occurs equally in $I=0$ and $I=1$.

\begin{table*}[placement specifier]
\begin{center}
{\tabcolsep = 2.7mm
\begin{tabular}{cccccccccc} \hline\hline
\multicolumn{2}{c}{$K_0$ [fm]}  &  \multicolumn{2}{c}{$K_1$ [fm]}  &  \multicolumn{2}{c}{$K_2$ [fm]}  &  \multicolumn{2}{c}{$K_3$ [fm]} \\ 
Re  &  Im  &  Re  &  Im  &  Re  &  Im  &  Re  &  Im   \\  
$-6.2261$  &  $-1.7382$  &  $-2.1909$  &  $-0.63300$  &  $-0.37668$  &  $-0.082052$  &  $-0.14782$  &  $-0.18206$  \\  \hline

\multicolumn{2}{c}{$K_4$ [fm]}   &  \multicolumn{2}{c}{$K_5$ [fm]}  &  \multicolumn{2}{c}{$K_6$ [fm]}  &  \multicolumn{2}{c}{$K_7$ [fm]}   \\ 
Re  &  Im  &  Re  &  Im  &  Re  &  Im  &  Re  &  Im   \\  
$2.9791$  &  $0.52183$  &  $ -0.53283$  &  $0.29745$  &  $-3.0760$  &  $-0.59389$  &  $1.5430$  &  $-0.062035$   \\  \hline
 
 \multicolumn{2}{c}{$K_8$ [fm]}  &  \multicolumn{2}{c}{$K_9$ [fm]}  &  \multicolumn{2}{c}{$K_{10}$ [fm]}  &  \multicolumn{2}{c}{}  \\ 
Re  &  Im  &  Re  &  Im  &  Re  &  Im  &  \multicolumn{2}{c}{}  \\  
$0.64668$  &  $0.36800$  &  $-0.59141$  &  $-0.17606$  &  $0.10746$  &  $0.026128$   &  \multicolumn{2}{c}{}  \\ \hline\hline
\end{tabular}  }
\caption{Coefficients $K_i$ in Eq.~\eqref{eq:fit} of the strength of SIDDHARTA potential $(I=1)$.}
\label{tab:KisidI1} 
\end{center}
\end{table*}
%

\subsection{Spacial structure of $\Lambda(1405)$}

We have succeeded in constructing the new $\bar{K}N$ local potential reliable for the quantitative calculations of the $\bar{K}$ nuclei. In this section, as a direct application of this new potential, we estimate the $\bar{K}N$ distance to understand the spacial structure of $\Lambda(1405)$. 

Generally, the unstable states are expressed as the poles of the scattering amplitude in the complex energy plane. As an analogy of a bound state, the spacial structure of an unstable state is reflected in the wave function at the pole energy.
With the solution of the radial Schr\"odinger equation $u_z(r)$ at the complex energy $z$, the $\bar{K}N$ wave function in $s$ wave is written as
\begin{align}
\psi_z(r) = \frac{1}{\sqrt{4\pi}}\frac{u_z(r)}{r}.  \label{eq:wavefcn}
\end{align}
The wave function of a resonance state diverges at $r\to \infty$. Hence the wave function cannot be normalized by the standard normalization,
\begin{align}
\int d{\bm r}\ {|\psi(r)|}^2 =1.  \label{eq:usualnor} 
\end{align}
Alternatively, the wave function of the non-Hermitian problem can be normalized with the Gamow vector  labeled by the index $G$ \cite{Hokkyo:1965,Berggren:1968zz},
\begin{align}
\int d{\bm r}\ {\psi_G(r)}^2 =1.  \label{eq:Gamownor} 
\end{align}
In the present problem, the poles of $\Lambda(1405)$ are in the physical Riemann sheet of the $\bar{K}N$ channel. Because the corresponding eigenmomentum has the positive imaginary part, the wave function converges at $r\to \infty$.\footnote{In the coupled-channel formulation, the wave function of the $\pi\Sigma$ channel diverges at $r\to \infty$.} Hence both the prescriptions \eqref{eq:usualnor} and \eqref{eq:Gamownor} are applicable (see also Appendix~A).

As explained in Appendix~B, for a problem with an energy-dependent potential, we should modify the normalization condition to ensure the conservation of the norm and the orthogonality relation between two states. The modified normalization condition is
\begin{align}
\int d{\bm r}\ \left( 1-\frac{\partial U(r,E)}{\partial E} \right){\psi^{\rm mod}_G(r)}^2 =1.  \label{eq:EdepGamownor} 
\end{align}
The expectation value of an operator should be modified in the similar way. For comparison, we calculate both the expectation values with Eq.~\eqref{eq:Gamownor} and Eq.~\eqref{eq:EdepGamownor}, and label the latter one by the index ``mod".

\begin{figure}[tb]
\begin{center}
\includegraphics[width=8cm,bb=0 0 834 591]{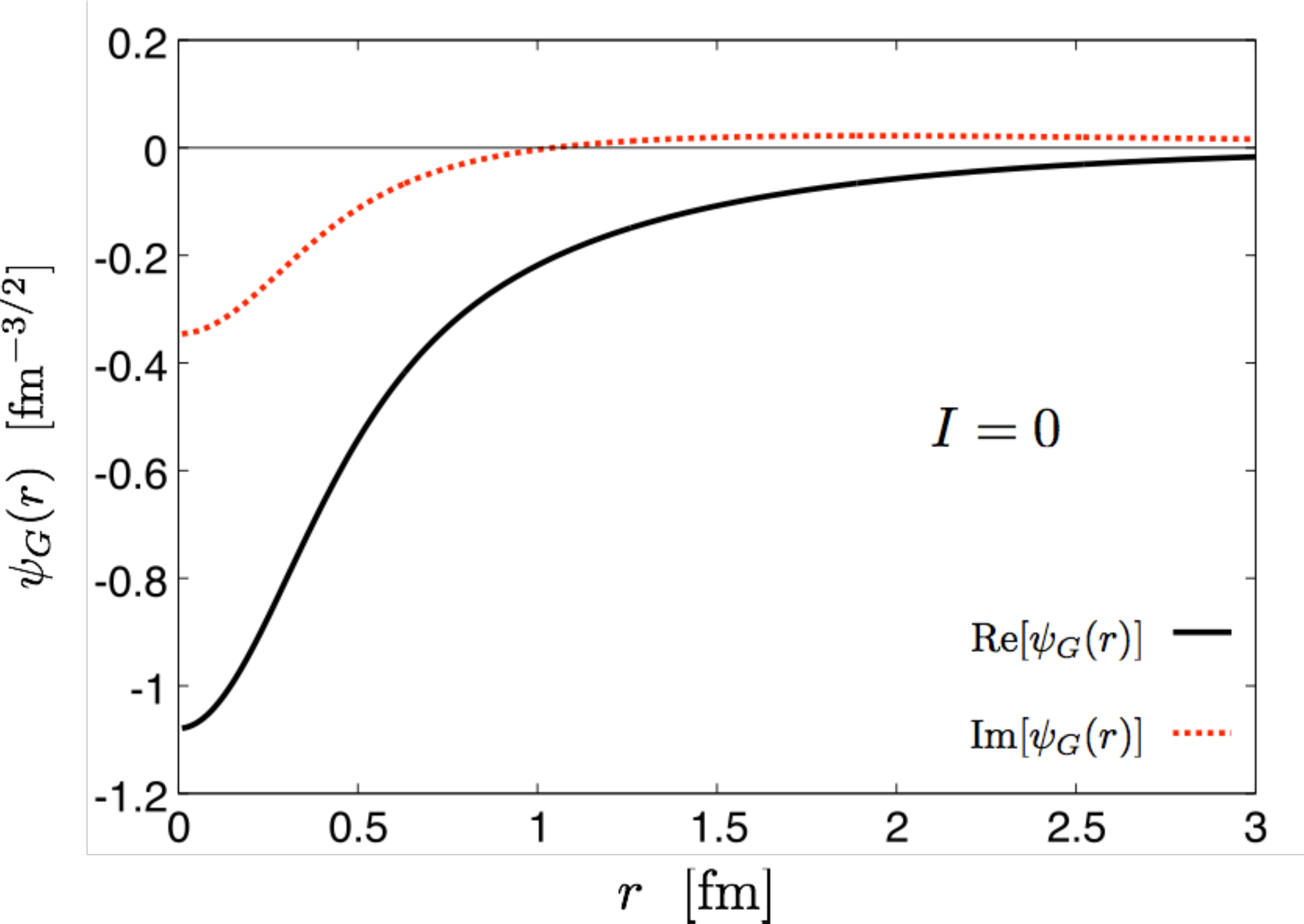}
\caption{(Color online) $I=0$ $\bar{K}N$ wave function $\psi_G$ with the normalization~\eqref{eq:Gamownor} at the $\Lambda(1405)$ pole energy, $1423.97-26.28i$ MeV. The real part is shown by the solid line, and the imaginary part is shown by the dotted line.}
\label{fig:psi}  
\end{center}
\end{figure}

The result of the wave function normalized with Eq.~\eqref{eq:Gamownor} is shown in Fig.~\ref{fig:psi}. Here we use the precise $\bar{K}N$ pole energy, $1423.97-26.28i\ {\rm MeV}$, in order to achieve the enough convergence at $r\sim 10\ {\rm fm}$. 
We note that the wave function has an imaginary part where the phase is uniquely determined by the normalization \eqref{eq:Gamownor}.
With this wave function, we calculate the expectation value of $r^2$ as
\begin{align}
\langle r^2 \rangle_G &\equiv \int d{\bm r}\ r^2{\psi_G(r)}^2.  \label{eq:Gamowexp}
\end{align}
The result of the mean squared distance (the root mean squared distance) of the antikaon and the nucleon is $\langle r^2 \rangle_G = 0.79-1.21i\ {\rm fm},\ (\sqrt{\langle r^2 \rangle_G} = 1.06-0.57i\ {\rm fm})$. 
Similarly, the distance with the modified normalization condition~\eqref{eq:EdepGamownor} can be calculated as
\begin{align}
\langle r^2 \rangle^{\rm mod}_G &\equiv \int d{\bm r}\ r^2\left( 1-\frac{\partial U(r,E)}{\partial E} \right){\psi^{\rm mod}_G(r)}^2.  \label{eq:EdepGamowexp}
\end{align}
The result of the mean squared distance (the root mean squared distance) is $\langle r^2 \rangle^{\rm mod}_G = 0.71-1.26i\ {\rm fm},\ (\sqrt{\langle r^2 \rangle^{\rm mod}_G} = 1.04-0.61i\ {\rm fm})$. It turns out that the modification of the normalization condition does not change the quantitative result very much.

In Table~\ref{tab:r^2}, we compare these results with the previous estimations. In Ref.~\cite{Sekihara:2012xp}, the radius is calculated by the form factor of $\Lambda(1405)$ in the chiral unitary model~\cite{Sekihara:2008qk,Sekihara:2010uz}. The result in Ref.~\cite{Dote:2014ema} is obtained by the $\bar{K}N$ wave function in the complex scaling method with the coupled-channel potential model. In both cases, the leading order Weinberg-Tomozawa interaction is used, without the constraint by the SIDDHARTA data. The present result from the next-to-leading order chiral interaction with the SIDDHARTA constraint quantitatively confirms the results of the previous works. 

\begin{table}[tb]
\begin{center}
{\tabcolsep = 3.5mm
\begin{tabular}{cc} \hline\hline
 & $\sqrt{\langle r^2\rangle_{G}}\ [{\rm fm}] $  \\ \hline 
SIDDHARTA potential  &  $1.06-0.57i$    \\ 
SIDDHARTA potential (modified)  &  $1.04-0.61i$    \\ 
Ref.~\cite{Sekihara:2012xp}  &  $1.22-0.63i$   \\
Ref.~\cite{Dote:2014ema}  & $1.22-0.47i$   \\ 
estimation from eigenmomentum  &  $0.85-0.58i$   \\ \hline\hline
\end{tabular}  }
\caption{Averaged $\bar{K}N$ distances at the $\Lambda(1405)$ energy. 
For comparison, we show the estimations from the form factor of $\Lambda(1405)$~\cite{Sekihara:2012xp} and  from the coupled channel potential model~\cite{Dote:2014ema}.}
\label{tab:r^2} 
\end{center}
\end{table}

If the spacial extent of the $\Lambda(1405)$ wave function is sufficiently larger than the potential range, the radius is determined mainly by the tail of the wave function.
Because the tail is related to the eigenenergy, we can estimate the spacial extent from the eigenenergy $E$ (see Appendix~A)
\begin{align}
\sqrt{\langle r^2 \rangle_{G}} \sim \frac{1}{\sqrt{2\kappa^2}} = \frac{1}{2\sqrt{-\mu E}}  \label{eq:r_zerorange}=0.85-0.58i\ {\rm fm} .
\end{align}
In this case, the value is same order as the $\sqrt{\langle r^2 \rangle_G}$ from the wave function. This means that the $\bar{K}N$ distance is sufficiently larger than the range of the potential. 

\begin{figure}[tb]
\begin{center}
\includegraphics[width=8cm,bb=0 0 760 542]{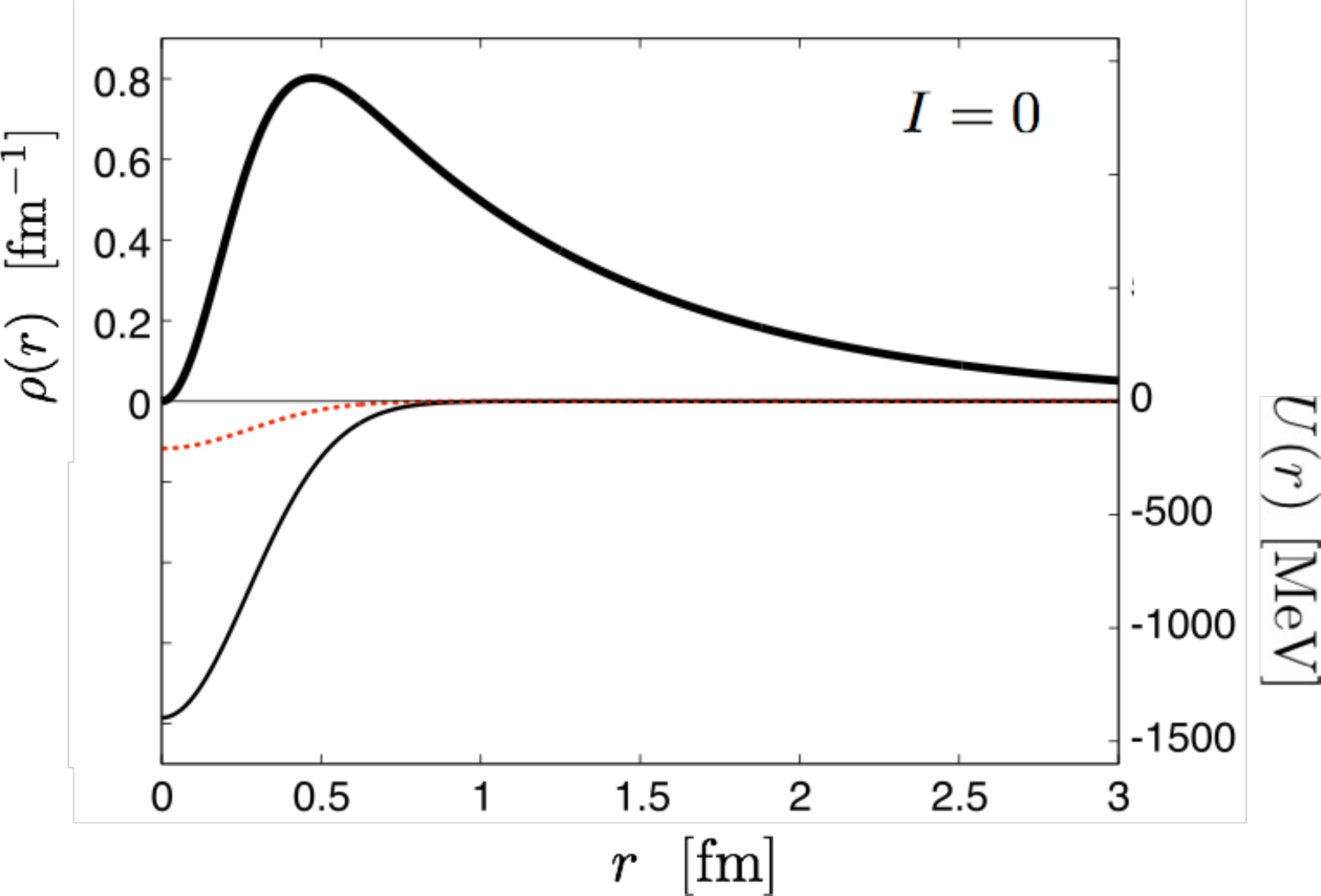}
\caption{(Color online) $\bar{K}N$ density distribution $\rho$ in Eq.~\eqref{eq:density} (thick solid line), the real part of SIDDHARTA potential $(I=0)$ (thin solid line) and the imaginary part (dotted line)
at the $\Lambda(1405)$ pole energy, $1423.97-26.28i$ MeV. }
\label{fig:density}  
\end{center}
\end{figure}

Though $\langle r^2 \rangle_G$ or $\langle r^2 \rangle^{\rm mod}_G$ give us some information about the spacial structure, it is not straightforward to interpret the complex number. As explained in Appendix~A, the dumpling of the wave function outside the potential range is related to the standard expectation value with normalization \eqref{eq:usualnor},
\begin{align}
\langle r^2 \rangle \equiv \int d{\bm r}\ r^2 |\psi(r)|^2.  \label{eq:usualexp}
\end{align}
Here we regard this quantity as the measure of the $\bar{K}N$ distance. 
As explained in Appendix~B, the modification of the norm due to the energy dependence of the potential cannot be applied without using the Gamow vector. Therefore, we calculate the $\bar{K}N$ distance with Eq.~\eqref{eq:usualexp}. This is convincing, because the values of $\sqrt{\langle r^2 \rangle_G}$ and $\sqrt{\langle r^2 \rangle^{\rm mod}_G}$ are almost same as shown in Table~\ref{tab:r^2}.
The result of the $\bar{K}N$ distance with Eq.~\eqref{eq:usualexp} is found to be
\begin{align}
\sqrt{\langle r^2 \rangle} = 1.44\ {\rm fm} .
\end{align}
Considering the charge radii of the proton and $K^-$ are about 0.85 fm and 0.55 fm \cite{Beringer:1900zz}, we find that the $\bar{K}N$ distance is relatively large in comparison with the usual hadron size. Therefore we conclude that $\Lambda(1405)$ is the molecular state of the antikaon and the nucleon. 
In order to visualize the spacial structure, we define the density distribution
\begin{align}
\rho(r) = r^2 {|\psi(r)|}^2,   \label{eq:density}
\end{align}
which is shown in Fig.~\ref{fig:density}.
The substantial distribution exists outside the potential range $b=0.38\ {\rm fm}$.

Finally, we investigate the $\bar{K}N$ distance against the change of the potential range $b$. The strength of the potential is adjusted to reproduce the original amplitude for each $b$.
The results of $\sqrt{\langle r^2 \rangle}_G$ and $\sqrt{\langle r^2 \rangle}$ are shown in Table \ref{tab:r^2_b}.
\begin{table}[tb]
\begin{center}
{\tabcolsep = 5mm
\begin{tabular}{lcc} \hline\hline
$b$\ [fm] & $\sqrt{\langle r^2\rangle}_G\ [{\rm fm}] $  &  $\sqrt{\langle r^2\rangle}\ [{\rm fm}] $  \\ \hline
0.2  &  $0.96-0.58i$  &  1.35  \\ 
0.38  &  $1.06-0.57i$  &  1.44  \\ 
0.4  &  $1.07-0.57i$  &  1.48  \\ 
0.6  & $1.18-0.57i$  &  1.57  \\ 
0.8  & $1.29-0.57i$  &  1.67  \\ \hline\hline
\end{tabular}  }
\caption{The results of the average $\bar{K}N$ distance with the Gamow vector normalization $\sqrt{\langle r^2 \rangle_G}$ and with the standard normalization method $\sqrt{\langle r^2 \rangle}$ against the change of the potential range $b$.}
\label{tab:r^2_b} 
\end{center}
\end{table}
In all cases, the values of $\sqrt{\langle r^2 \rangle}$ remain larger than the typical hadron size. We find the qualitative picture of the molecular state is valid irrespective of the potential range.

\section{Summary}  \label{sec:summary}  

We have constructed the new $\bar{K}N$ local potential (SIDDHARTA potential) which reproduces the scattering amplitude from the chiral SU(3) dynamics. 
In the construction procedure, we have paid attention to the two steps: the precision in the complex energy plane and the constraint from the recent SIDDHARTA data. This new potential is useful for the quantitative calculation of the interesting systems such as $\bar{K}$ few-body systems and the $\Lambda(1405)$. 

We first establish the procedure of potential construction by improving the previous work~\cite{Hyodo:2007jq}. The previous potential almost reproduced the original amplitude on the real energy axis, while we have found that there is a substantial deviation in the complex energy plane, including the poles of $\Lambda(1405)$. Therefore we need to improve the potential construction procedure to reproduce the original amplitude even in the complex energy plane. 
We find that the reduction of the deviation on the real energy axis $\Delta F_{\rm real}$ in the wide parametrized range is important, based on the uniqueness of the analytic continuation  in the complex energy plane. 
Thanks to these improvements, we have succeeded in reproducing the original amplitude in the drastically large region in the complex plane including the  two poles of $\Lambda(1405)$. 

Next, we have applied the new procedure to the amplitude with the SIDDHARTA constraint to construct the realistic $\bar{K}N$ potential. Here we produce the $\bar{K}N$ amplitude with isospin symmetry from the coupled-channel chiral model in Refs.~\cite{Ikeda:2011pi,Ikeda:2012au}. 
Based on these amplitudes, we have constructed the realistic $\bar{K}N$ local potentials for the $I=0$ and $I=1$ channels. The $I=0$ local potential reproduces the original amplitude in the complex energy plane including the poles of $\Lambda(1405)$. At the present time, this is the most reliable local $\bar{K}N$ potential for the quantitative calculations.  
Applying this new potential to $\Lambda(1405)$, we have estimated the spacial structure of $\Lambda(1405)$. The mean distance of $\bar{K}$ and $N$ is found to be 1.44 fm. This result shows the meson-baryon molecular nature of $\Lambda(1405)$. 

As a future perspective, the calculation of the $\bar{K}NN$ system with the new reliable potential is of particular importance. We hope that this result will bring new insight in the theoretical and experimental studies of the $\bar{K}NN$ system.

\section*{Acknowledgments}

The authors thank Yoshiko Kanada-En'yo for useful discussion on the mean squared radius of the quasi-bound state. This work is supported in part by JSPS KAKENHI Grants No. 24740152 and by the Yukawa International Program for Quark-Hadron Sciences (YIPQS).

\appendix
\section*{Appendix~A : average distance of quasi-bound state}

In this Appendix, we consider the mean squared radius of a quasi-bound state in comparison with a bound state in the zero range limit. When the spatial extent of the wave function is much larger than the potential range, we can treat the potential in the zero range approximation. In this case, the mean squared radius $\langle r^2 \rangle$ is determined only by the eigenmomentum $k$, where $k=i\kappa\ (\kappa>0) $ for the bound state and $k=i\kappa-\gamma\ (\kappa,\gamma>0)$  for the quasi-bound state.
 
In the zero range limit, the radial wave function of the (quasi-)bound state in $s$ wave is written as
 \begin{align}
u(r) &\to A e^{ikr},   \label{eq:wavefcn_zero}
 \end{align}
where $u(r)$ is related to the wave function as $\psi(r) = u(r)/(\sqrt{4\pi}r)$.
The normalization condition determines the factor $A$.  We consider the following two normalization conditions,
\begin{align}
\langle \psi|\psi\rangle &=\int d\bm{r}\ |\psi(r)|^2=1 ,\label{eq:nor_usual}  \\
_G\langle \psi|\psi\rangle_G &=\int d\bm{r}\ \psi(r)^2=1 ,\label{eq:nor_Gamow}
\end{align}
where the former is the standard normalization whereas the latter uses the Gamow vector labeled by $G$.
In the zero range limit, these conditions are expressed as
\begin{align}
\langle \psi|\psi\rangle &\to |A|^2\int_0^\infty dr\ e^{-2{\rm Im}[k] r} =\frac{|A|^2}{2{\rm Im}[k]}=1,  \notag \\
_G\langle \psi|\psi\rangle_G &\to A^2\int_0^\infty dr\ e^{2ik r}=\frac{A^2}{2ik}=1.  \notag
\end{align}
These integrals converge for the bound state or the quasi-bound state because ${\rm Im}[k]>0$ \footnote{The resonance wave function (Im[$\kappa$]$<0$) can be normalized only by Eq.~\eqref{eq:nor_Gamow} with appropriate prescription \cite{Berggren:1968zz}.}. The normalized wave functions are written as
\begin{align}
\psi(r) &\to \frac{A}{|A|}\sqrt{\frac{{\rm Im}[k]}{2\pi}}\frac{e^{ikr}}{r} \equiv e^{i\theta} \sqrt{\frac{{\rm Im}[k]}{2\pi}}\frac{e^{ikr}}{r},  \label{eq:wavefcn_nor_usual} \\
\psi_G(r) &\to \sqrt{\frac{-ik}{2\pi}}\frac{e^{ikr}}{r},  \label{eq:wavefcn_nor_Gamow}
\end{align}
where $\theta$ is an arbitrary real constant. In the standard normalization, physical observables are independent of the phase of the wave function, so $\theta$ is an irrelevant phase. In the case of the bound state $k=i\kappa$, Eq.~\eqref{eq:wavefcn_nor_usual} and Eq.~\eqref{eq:wavefcn_nor_Gamow} are equivalent. On the other hand, these wave functions of the quasi-bound state are in general different, $\psi\neq\psi_G$.

With these wave functions, we can calculate the mean squared radius,
\begin{align}
\langle r^2 \rangle &=\int d\bm{r}\ r^2|\psi(r)|^2 \notag \\
&\to \frac{{\rm Im}k}{2\pi}4\pi\int_0^\infty dr\ r^2e^{-2{\rm Im}[k] r} \notag \\
&=\frac{1}{2{(\rm Im}[k])^2}, \label{eq:r2_usual}  \\
\langle r^2 \rangle_G &=\int d\bm{r}\ r^2\psi_G(r)^2 \notag \\
&\to \frac{-ik}{2\pi}4\pi\int_0^\infty dr\ r^2e^{2ik r} \notag \\
&=\frac{1}{2(-ik)^2}. \label{eq:r2_Gamow}
\end{align}
In the case of the bound state $k=i\kappa\ (\kappa>0)$, both the normalizations give the same result,
\begin{align}
\langle r^2 \rangle = \langle r^2 \rangle_G = \frac{1}{2\kappa^2}\quad \text{(bound state)}.  \label{eq:r2_bound}
\end{align}
For the quasi-bound state, Eq.~\eqref{eq:r2_usual} and Eq.~\eqref{eq:r2_Gamow} are different: 
\begin{align}
\langle r^2 \rangle &=  \frac{1}{2\kappa^2},   \label{eq:r2_quasi}\\
\langle r^2 \rangle_G &=  \frac{1}{2\kappa^2+4i\kappa\gamma -2\gamma^2}\quad \text{(quasibound state)}. \label{eq:r2_quasigamow}
\end{align}
Eq.~\eqref{eq:r2_quasi} gives a real number for the mean squared radius, while Eq.~\eqref{eq:r2_quasigamow} gives a complex number.
It is common to use the Gamow vector \cite{Hokkyo:1965,Berggren:1968zz,delaMadrid:2002cz} for unstable states. Because the radial wave function of the quasi-bound state asymptotically behaves as $e^{ikr}$, Eq.~\eqref{eq:r2_Gamow} is the natural extension of the bound state. We therefore use the normalization with the Gamow vector in Fig.~\ref{fig:psi} and Table~\ref{tab:r^2}. 

On the other hand, it is difficult to extract the spacial information from the complex $\langle r^2 \rangle_G$ in Eq.~\eqref{eq:r2_quasigamow}. We note that the dumping of the wave function of the quasi-bound state is expressed by $e^{-{\rm Im}[k]r}$ in the asymptotic behavior $e^{ikr}$.
In this sense, we consider that the real $\langle r^2\rangle$ with the standard normalization, which is determined by ${\rm Im}[k]$, can be interpreted as the spacial extent of the quasi-bound state.
Hence, in this paper, we use $\langle r^2\rangle$ to estimate the spacial extent of the $\bar{K}N$ quasi-bound state, $\Lambda(1405)$.

\section*{Appendix~B : Energy dependent complex potential}

As explained in Refs.~\cite{Lepage:1997cs,Lepage:1977gd,Rescigno:1986,Sazdjian:1986qn,Formanek:2003zc,Miloslav:2003}, the careful treatment is necessary for the system with the energy-dependent potential. Here, we explain the treatment in the cases of real potential and complex one. 

First, we summarize the case of the energy-dependent real potential, following Ref.~\cite{Formanek:2003zc}. We consider the Schr\"odinger equation with a time-dependent wave function $\Psi({\bm r},t)$
\footnote{Here, we assume that the wave function is normalizable.} with $\mu=1$,
\begin{align}
i\frac{\partial \Psi({\bm r},t)}{\partial t} &= H\Psi({\bm r},t)  \notag \\
&=\left[-\frac{1}{2}\nabla^2 + V({\bm r},i\frac{\partial}{\partial t}) \right]\Psi({\bm r},t). \label{eq:tdepschro}
\end{align}
For an eigenfunction of the Hamiltonian, $\Psi_E({\bm r},t)=e^{-iEt}\psi_E({\bm r})$, the time-independent Schr\"odinger equation becomes
\begin{align}
H \psi_E({\bm r})&= \left[ -\frac{1}{2}\nabla^2 + V({\bm r},E) \right] \psi_E({\bm r}) = E\psi_E({\bm r}).  \label{eq:tindepschro}
\end{align}
With Eq.~\eqref{eq:tdepschro}, the continuity equation for energy-dependent potential can be calculated as follows,
\begin{align}
\frac{\partial}{\partial t}P &= \frac{\partial \Psi_{E^\prime}^*}{\partial t}\Psi_E + \Psi_{E^\prime}^*\frac{\partial \Psi_E}{\partial t}  \notag \\ 
&=\left[-i\left\{ -\frac{1}{2}\nabla^2+V({\bm r},E^\prime)\right\} \Psi_{E^\prime} \right]^*\Psi_E \notag \\
&\phantom{aaaaa}+ \Psi_{E^\prime}^*\left[-i\left\{ -\frac{1}{2}\nabla^2+V({\bm r},E)\right\} \Psi_E \right]  \notag \\
&=-\nabla\cdot {\bm j} + i\Psi_{E^\prime}^* \big[V({\bm r},E^\prime)-V({\bm r},E)\big] \Psi_E,  \label{eq:cont}
\end{align}
where 
\begin{align}
P&=\Psi_{E^\prime}^*({\bm r},t)\Psi_E({\bm r},t),\notag \\
{\bm j}&=-\frac{i}{2}\left[ \Psi_{E^\prime}^*({\bm r},t)\nabla\Psi_E({\bm r},t)-\big\{ \nabla\Psi_{E^\prime}^*({\bm r},t)\big\} \Psi_E({\bm r},t)   \right].  \notag
\end{align}
For the energy-independent potential, the second term of the last line in Eq.~\eqref{eq:cont} disappears and the usual continuity equation $\partial P/\partial t = -\nabla\cdot{\bm j}$ can be hold. However, for the energy-dependent potential, the additional term have to be included. Using the Schr\"odinger equation, $i\partial \Psi_E/{\partial t}=E\Psi_E$, we obtain the relation,
\begin{align}
\frac{\partial}{\partial t}&\left\{ \Psi_{E^\prime}^* \left[ \frac{V(E^\prime)-V(E)}{E^\prime-E} \right]\Psi_E \right\} \notag \\
&= \frac{\partial \Psi_{E^\prime}^*}{\partial t} \left[\frac{V(E^\prime)-V(E)}{E^\prime-E}  \right]\Psi_E \notag \\
&\phantom{aaaaa}+ \Psi_{E^\prime}^*\left[\frac{V(E^\prime)-V(E)}{E^\prime-E}  \right]\frac{\partial \Psi_E}{\partial t}  \notag \\
&=\left\{-iE^\prime \Psi_{E^\prime} \right\}^* \left[ \frac{V(E^\prime)-V(E)}{E^\prime-E} \right] \Psi_E \notag \\
&\phantom{aaaaa}+ \Psi_{E^\prime}^* \left[ \frac{V(E^\prime)-V(E)}{E^\prime-E} \right] \left\{-iE^\prime \Psi_E \right\}  \notag \\
&= i\Psi_{E^\prime}^*\left[ V(E^\prime)-V(E) \right] \Psi_E,  \label{eq:addterm}
\end{align}
and the continuity equation for the energy-dependent potential can be modified as 
\begin{align}
\frac{\partial}{\partial t}(P+P_a) = -\nabla\cdot{\bm j},  \label{eq:edepcont}
\end{align}
where
\begin{align}
P_a = -\Psi_{E^\prime}^*({\bm r},t) \left[ \frac{V({\bm r},E^\prime)-V({\bm r},E)}{E^\prime-E} \right]\Psi_E({\bm r},t).  \notag
\end{align}
Therefore, taking the limit of $E^\prime\to E$, the norm $N$ can be modified as 
\begin{align}
N &= \int d{\bm r}\ \Psi_E^*({\bm r},t) \left[1-\frac{\partial V({\bm r},E)}{\partial E} \right] \Psi_E({\bm r},t)  \notag \\
&= \int d{\bm r}\ \psi_E^*({\bm r}) \left[1-\frac{\partial V({\bm r},E)}{\partial E} \right] \psi_E({\bm r}).  \label{eq:norm}
\end{align}
Furthermore, the orthogonality relation can be modified as
\begin{align}
\int d{\bm r}\ \psi_{E^\prime}^*({\bm r}) \left[1-\frac{V({\bm r},E^\prime)-V({\bm r},E)}{E^\prime-E} \right] &\psi_E({\bm r}) = 0,  \label{eq:ortho}  \\
&(E^\prime \ne E).  \notag
\end{align}
Actually, the usual orthogonality relation is not satisfied because the term with $P_a$ remains nonzero after integrating  Eq.~\eqref{eq:edepcont} with respect to ${\bm r}$. 

The above method cannot be directly applied to the case of the complex potential. In this case, following the same procedure, $P_a$ is obtained as 
\begin{align}
P_a = -\Psi_{E^\prime}^*({\bm r},t) \left[ \frac{V^*({\bm r},E^\prime)-V({\bm r},E)}{E^\prime-E} \right]\Psi_E({\bm r},t).  \notag
\end{align}
Because $V^*(E)\ne V(E)$ for the complex potential, taking the limit of $E^\prime\to E$, this term diverges and does not become the derivative of the potential. 
When we treat the complex potential problem, it is common to use the adjoint wave function $\Psi^\dag$~\cite{Hokkyo:1965}, which satisfies the Schr\"odinger equation
\begin{align}
i\frac{\partial \Psi^\dag({\bm r},t)}{\partial t} = H^\dag \Psi^\dag({\bm r},t).  \label{eq:adj}
\end{align}
Under the adequate boundary condition~\cite{Hokkyo:1965},  the time-independent eigenfunction $\psi_E^\dag$ have the following properties,
\begin{align}
H^\dag\psi^\dag_E({\bm r}) &= H^*\psi^\dag_E({\bm r})  \notag \\
&=\left( -\frac{1}{2m}\nabla^2+V^*({\bm r},E)  \right) \psi^\dag_E({\bm r}) = E^* \psi^\dag_E({\bm r}),  \label{eq:adjsch} \\
\psi_E^\dag({\bm r}) &= \psi_E^*({\bm r}),  \label{eq:adjrel}  \\
\Psi_E^\dag({\bm r},t) &= e^{-iE^*t}\psi_E^\dag({\bm r}) = e^{-iE^*t}\psi_E^*({\bm r}).  \notag 
\end{align}
With this adjoint wave function, we consider the continuity equation again, labeling the quantities with the adjoint function by the index ``$G$".
\begin{align}
\frac{\partial}{\partial t}P^G &= \frac{\partial \Psi_{E^\prime}^{\dag *}}{\partial t}\Psi_E + \Psi_{E^\prime}^{\dag *}\frac{\partial \Psi_E}{\partial t}  \notag \\ 
&=\left[-i\left\{ -\frac{1}{2}\nabla^2+V^*(E^\prime)\right\} \Psi^\dag_{E^\prime} \right]^*\Psi_E \notag \\
&\phantom{aaaaa}+ \Psi_{E^\prime}^{\dag *}\left[-i\left\{ -\frac{1}{2}\nabla^2+V(E)\right\} \Psi_E \right]  \notag \\
&=-\nabla\cdot{\bm j}^G + i\Psi_{E^\prime}^{\dag*} \left(V(E^\prime)-V(E)\right) \Psi_E,  \label{eq:contgamow}
\end{align}
where 
\begin{align}
P^G&=\Psi_{E^\prime}^{\dag *}({\bm r},t)\Psi_E({\bm r},t),\notag \\
{\bm j}^G&=-\frac{i}{2}\left[ \Psi_{E^\prime}^{\dag *}({\bm r},t)\nabla\Psi_E({\bm r},t)- \big\{ \nabla\Psi_{E^\prime}^{\dag *}({\bm r},t) \big\} \Psi_E({\bm r},t)   \right].  \notag
\end{align}
Following the same procedure as in Eq.~\eqref{eq:addterm}, the continuity equation for the complex energy-dependent potential can be satisfied.
\begin{align}
\frac{\partial}{\partial t}(P^G+P^G_a) = -\nabla\cdot{\bm j}^G,  \label{eq:edepcompcont}
\end{align}
where
\begin{align}
P^G_a = -\Psi_{E^\prime}^{\dag *}({\bm r},t) \left[ \frac{V({\bm r},E^\prime)-V({\bm r},E)}{E^\prime-E} \right]\Psi_E({\bm r},t).  \notag
\end{align}
Therefore, we should modify the norm and the orthogonality relation as
\begin{align}
N &= \int d{\bm r}\ \Psi_E^{\dag *}({\bm r},t) \left[1-\frac{\partial V({\bm r},E)}{\partial E} \right] \Psi_E({\bm r},t)  \notag \\
&= \int d{\bm r}\ \psi_E({\bm r}) \left[1-\frac{\partial V({\bm r},E)}{\partial E} \right] \psi_E({\bm r}),  \label{eq:norm}
\end{align}
\begin{align}
\int d{\bm r}\ \psi_{E^\prime}({\bm r}) \left[1-\frac{V({\bm r},E^\prime)-V({\bm r},E)}{E^\prime-E} \right] &\psi_E({\bm r}) = 0,  \label{eq:ortho}  \\
&(E^\prime \ne E).  \notag
\end{align}


%


\end{document}